\documentclass[twocolumn,secnumarabic,amssymb, nobibnotes, aps, prl, superscriptaddress]{revtex4-1}

\usepackage{float}
\usepackage{graphicx}
\usepackage{subfigure}
\usepackage{amsmath} 
\usepackage{epstopdf}
\usepackage{setspace}
\usepackage[shortlabels]{enumitem}
\usepackage{mathrsfs}
\usepackage{subfig}
\usepackage{color,soul}
\usepackage[dvipsnames]{xcolor}
\usepackage{dcolumn,amssymb,subfig}
\usepackage{bm}
\usepackage{ulem}

\renewcommand{\vec}[1]{\bm{#1}}
\newcommand{\tens}[1]{\mathbf{#1}}
\newcommand{\bnabla}{\vec{\nabla}}

\newcommand{\fig}[1]{\textbf{Fig.\ref{#1}}}

\begin{document}

\title{Binding self-propelled topological defects in active turbulence}

\author{Kristian Thijssen}
\affiliation{The Rudolf Peierls Centre for Theoretical Physics, Clarendon Laboratory, Parks Road, Oxford, OX1 3PU, UK.}

\author{Amin Doostmohammadi}
\affiliation{Niels Bohr Institute, University of Copenhagen, Blegdamsvej 17, 2100 Copenhagen, Denmark}
\begin{abstract}
We report on the emergence of stable self-propelled bound defects in monolayers of active nematics, which form virtual full-integer topological defects in the form of vortices and asters. Through numerical simulations and analytical arguments, we identify the phase-space of the bound defect formation in active nematic monolayers. It is shown that an intricate synergy between the nature of active stresses and the flow-aligning behaviour of active particles can stabilise the motion of self-propelled positive half-integer defects into specific bound structures. Our findings uncover new complexities in active nematics with potential for triggering new experiments and theories.
\end{abstract}
\maketitle

Whether they are fluxons in a superconductor \cite{mclaughlin1978perturbation,sakai1993fluxons,wambaugh1999superconducting}, vortices in a superfluid \cite{salomaa1987quantized,schwarz1988three,zwierlein2005vortices}, or disclinations in a liquid crystal \cite{DeGennes}, topological defects mark the breakdown of an order parameter in the system.
Though they are in principle imperfections, topological defects have emerged as indispensable features of matter for various applications from skyrmions in quantum field theory~\cite{fert2013skyrmions,nagaosa2013topological} to blue phases in liquid crystals~\cite{wright1989crystalline}.
Intriguingly, the functional role of topological defects has also been recently identified in a growing number of biological processes.
Striking examples are cytoskeletal topological defects determining growth axis of animal Hydra~\cite{livshits2017structural,maroudas2020topological}, defects in cell orientation governing cell death and extrusion in epithelial tissues~\cite{Saw2017}, defects in bacterial biofilms leading to layer formation~\cite{yaman2019}, 
and defects as local hotspots of mound formation in neural stem cells~\cite{kawaguchi2017topological}.

Despite having similar topologically protected structures, as no local rearrangement of the order parameter can remove them \cite{DeGennes}, the defects in these living biological materials show distinct dynamical behaviour compared to their non-living counterparts \cite{doostmohammadi2018active}.
This is due to the continuous injection of energy at the local level of each individual living particle~\cite{ramaswamy2010mechanics,Marchetti13,Bechinger2016,julicher2018hydrodynamic}. Every particle extracts energy from the surrounding medium and converts it into mechanical work in the form of self-propulsion and active stress generation.
As a result of this activity topological defects with broken symmetry, $+1/2$ defects, can self-propel in active materials \cite{Sanchez2012,giomi2014defect,kumar2018tunable,Li19}. 
  \begin{figure}[h] 
    \centering
    \includegraphics[width=0.5\textwidth]{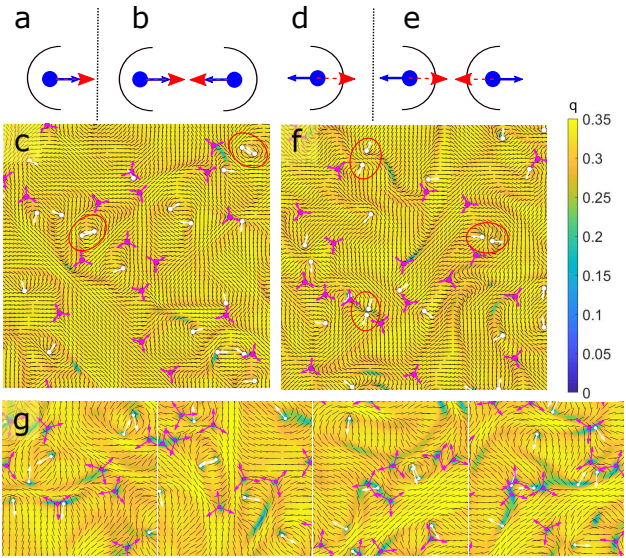}
    \caption{
    Emergence of bound defect pairs.
    (a,d) Schematic movement of a $+1/2$ topological defect with polar, head-to-tail, axis ({where blue arrows denotes the tail direction}) and a self-propulsion direction (red arrows) for a contractile (a) or an extensile system (d) respectively.
    (b,e) Two $+1/2$ defects move toward each other and form a bound $+1$ defect with a vortex-like structure for contractile (b) or an aster-like structure for extensile (e) systems. 
    (c,f) Snapshots of (c) a contractile system with positive flow-aligning parameter $\nu=1.2$ or (f) an extensile system with negative flow-aligning parameter $\nu=-1.2$. The colormap illustrates the nematic ordering $q$ and $+1/2$, $-1/2$ defects are highlighted by comets and trefoils, respectively. {Snapshots of bound defect dynamics over time (see Movie 1).}
}
\label{fig:bound_defects}
\end{figure}

This self-propulsion of $+1/2$ defects leads to novel emergent behaviours, which have no equivalent in equilibrium systems. For example, as a consequence of the activity, a $+1/2$ defect experiences a torque that tends to align the defect in the direction of forces acting on it, leading to the unbinding of $\pm 1/2$ defects upon nucleation and active-nematic-to-isotropic phase transition~\cite{shankar2018defect}. 
Furthermore, experimental and theoretical studies show the emergence of long-ranged nematic~\cite{decamp2015orientational,thijssen2020large,pearce2020scale} and polar~\cite{putzig2016instabilities,shankar2019hydrodynamics} orientational alignment of self-propelled $+1/2$ defects that, at least theoretically, could even self-organise into positionally ordered lattice structures~\cite{doostmohammadi2016stabilization,oza2016antipolar,thijssen2020large}. 
Moreover, the strong flow fields generated by self-propelled $+1/2$ defects are shown to drive the formation of protrusions at free surfaces~\cite{doostmohammadi2016defect,metselaar2019topology}, dictating the morphology of active multiphase systems such as bacterial colonies~\cite{doostmohammadi2016defect,dell2018growing} and epithelial monolayers~\cite{morris2019active}.
Here, we report yet another new feature of active defects: self-organisation into stable like-charged bound pairs in the form of virtual full-integer defects - which has no parallel in equilibrium systems - and could be important in the formation of dynamically ordered stable structures in active systems. 
\begin{figure}[b]
    \centering
    \includegraphics[width=1.0\linewidth]{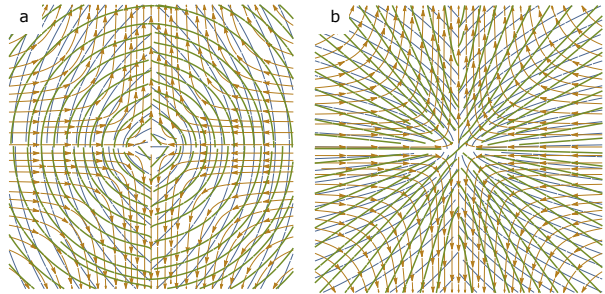}
    \caption{
    Analytical prediction of director reconfiguration in response to active flows of $+1/2$ defect pairs.
    (a) The contractile vortex-like configuration for $\nu\rightarrow 1$. (b) The extensile aster-like configuration for $\nu\rightarrow -1$.
    Blue dotted lines indicate the director field and their associated flow fields are illustrated with the orange streamlines.
    The green dotted lines indicate the expected reconfiguration of the director field in response to the active flows.
}
\label{fig:Binding_mech}
\end{figure}

While {\it half-integer} defects are prolific in active nematics \cite{doostmohammadi2018active}, the understanding of {\it full} integer defects remains obscure.
In fact, in two-dimensional layers of particles with nematic symmetry and in the absence of any specified anchoring at the boundaries, simple energy calculation shows that two half-integer defects have less elastic energy than one full-integer defect and as such any full integer defect should be inherently unstable~\cite{meyer1973existence}. 
Here, however, we show that in active nematics, self-propelled $+1/2$ defects can come together to form virtual full-integer defects in the form of asters and vortices. Importantly, this behaviour emerges in the bulk of the system in active turbulence regime \cite{Wensink2012} and in the absence of any confinement or any imposed anchoring boundary conditions.
We show that in addition to activity, the flow-aligning behaviour of active particles is a determining factor in the binding of two $+1/2$ defects into stable aster- and vortex-like structures.
This is important since the formation of such full-integer topological defects in 2D has until now only been associated with systems with polar symmetry~\cite{lee2001macroscopic,kruse2004asters,husain2017emergent,endresen2019topological}, confined systems subject to strong anchoring conditions~\cite{duclos2017topological,opathalage2019self,guillamat2020integer,turiv2020topology}  or externally applied stresses~\cite{rivas2019driven}.

To investigate the formation of full integer defects, we solve the full continuum equations of active nematohydrodynamics. 
This is commonly used in describing the spatio-temporal dynamics of a wide range of active systems, from microtubule-kinesin motor protein mixtures~\cite{Sanchez2012,hardouin2019reconfigurable} and actin filaments powered by myosin motors~\cite{zhang2018interplay}, to bacterial colonies~\cite{volfson2008biomechanical, zhang2010collective,dell2018growing,beroz2018verticalization} and dense assemblies of fibroblast cells~\cite{duclos2014perfect}. 
Within this framework, the dynamics of the active system is governed through the coupling between the evolution of a nematic tensor $\tens{Q}$ as the orientational order parameter and the velocity vector $\vec{u}$ as the slow variable.
The velocity field is determined from the generalized Stokes equation that comprises viscous, elastic, and active stresses (see  \cite{suppmat} for the full description of stress terms). 
Of particular importance here is the active stress term $\boldsymbol{\sigma}^{\text{act}}=-\zeta\tens{Q}$, which implies that any gradients in the nematic ordering generate flows~\cite{ramaswamy2010mechanics}. 
The sign of the activity coefficient $\zeta$ distinguishes between two types of active particles with $\zeta>0$ ($\zeta<0$) corresponding to extensile (contractile) active entities~\cite{ramaswamy2010mechanics}. 
The Beris-Edwards equation~\cite{BerisBook} governs the dynamics of the nematic tensor $\tens{Q}$: 
\begin{equation}
\partial_t\tens{Q}+\vec{u}\cdot\vec{\nabla}\tens{Q}-\tens{S}=\Gamma\tens{H},\nonumber
\end{equation}
with $\Gamma$ the rotational diffusion controlling the relaxation of the orientation through the molecular field $\tens{H}$ (see  \cite{suppmat} for the full form of the free energy and the molecular field). 
In addition to the advection with the flow, the co-rotational term $\tens{S} = \left(\lambda \tens{E}+\tens{\Omega}\right)\left(\tens{Q}+\frac{1}{3}\tens{I}\right) + \left(\tens{Q}+\frac{1}{3}\tens{I}\right)\left(\lambda \tens{E}-\tens{\Omega}\right)-2\lambda\left(\tens{Q}+\frac{1}{3}\tens{I}\right)\text{tr}\left(\tens{Q}:\bnabla\vec{u}\right)$ determines the response of the orientation field to the rate of strain $\tens{E}$ and the vorticity $\tens{\Omega}$ tensors. 
The nature of this response to flow gradients is controlled by the flow-aligning parameter $\lambda$: when the normalized flow-aligning parameter $|\nu|=\frac{|\lambda| 9q}{3q+4}>1$ (with $q$ denoting the magnitude of nematic order), the director aligns with an angle to the principal axis of flow deformation (See \cite{suppmat}). When $|\nu|<1$ the director tumbles in the vorticity field set by the flow gradients. 
The value of $\nu$ depends on the size, aspect ratio, and also interactions between the elongated particles~\cite{DeGennes,aigouy2010cell,duclos2018spontaneous}. 
For larger values of the flow-aligning parameter (in the flow-aligning regime), positive values implicate alignment parallel to the pure shear, while negative values describe particles that align perpendicular to the pure shear.  These different responses of the orientation field to flow gradients become particularly complex in the case of active materials since the flows are self-generated at the level of constituent elements that in turn respond to their actively self-induced flows.
As such, the exact mapping of the flow-aligning parameter in various experiments remains to be determined, but existing studies suggest negative values of $\lambda$ for  epithelial cells  forming Drosphila wing  \cite{aigouy2010cell}.

While the pivotal role of activity in setting the motion and ordering of self-propelled topological defect is well established both in models~\cite{giomi2014defect,shankar2018defect,shankar2019hydrodynamics,thijssen2020large} and experiments~\cite{decamp2015orientational,guillamat2017taming,hardouin2019reconfigurable,pearce2020scale}, the importance of the flow-aligning parameter and its impact on topological defects motion have been largely overlooked.
To this end, we begin by probing the behaviour of topological defects in the activity flow-aligning parameter ($\zeta-\nu$) phase space.  
We choose the model parameters in the range that has proven successful in describing spatio-temporal dynamics of microtubule/motor protein mixtures~\cite{hardouin2019reconfigurable} (see \cite{suppmat} for the simulation details). 
~Periodic boundary conditions are applied on all sides of the domain and simulations begin with zero velocities and random director field.

First, we consider the case of a positive flow-aligning parameter in the flow aligning regime with $\lambda=0.7$ ($\nu=1.2$). 
Since most of the numerical studies so far have only focused on extensile activities~\cite{srivastava2016negative,doostmohammadi2017onset,shendruk2017dancing,shankar2018defect,shankar2019hydrodynamics,doostmohammadi2018active,kumar2018tunable,santhosh2020activity,chandragiri2019active}, we begin by exploring the effect of contractile activity to shed light on possible new patterns. 
At low contractile activities, splay instabilities are followed by unbinding of $\pm 1/2$ defect pairs, and active turbulence is established, a process that is consistent with both previous theoretical predictions~\cite{ramaswamy2010mechanics,shankar2018defect} and numerical simulations~\cite{giomi2013defect,thampi2014instabilities}. However, by increasing the contractile activity in this flow-aligning regime, we observe that after $\pm 1/2$ unbinding events, pairs of $+1/2$ defects join together along their comet-shaped tails (\fig{fig:bound_defects}a), forming stable structures in the form of vortex-like topological defects (\fig{fig:bound_defects}b,c). 
Further increase in activity restores unbinding and active turbulence, destroying bound $+1/2$ defect pairs. 
Intriguingly, we do not observe similar activity-dependent $+1/2$ defect binding for a flow-tumbling regime, indicating that the underlying mechanism of bound $+1/2$ defect formation must be controlled by the flow-aligning behaviour of the director field.
\begin{figure}[t] 
    \centering
    \includegraphics[width=0.9\columnwidth]{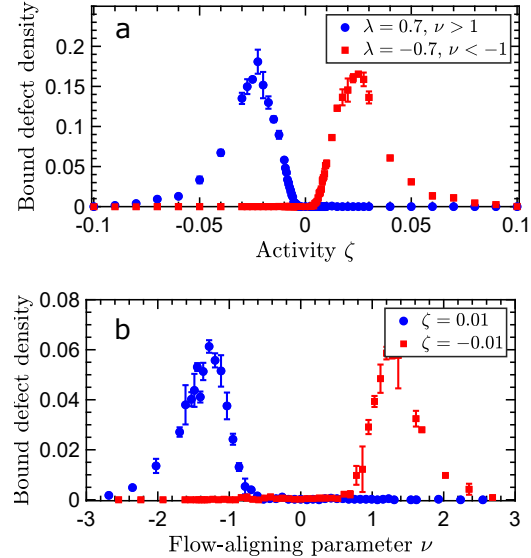}
    \caption{
    Numerical measurements of the bound defect density as a function of (a) activity and (b) flow-aligning parameter.
}
\label{fig:bound_Defects_time}
\end{figure}

To understand the mechanism of bound $+1/2$ defect pair formation, we consider the response of the director field associated with two interacting $+1/2$ topological defects to their corresponding activity-induced flow fields (\fig{fig:Binding_mech}a).
To this end, we consider a simplified setup, where only two $+1/2$ defects are near each other in a vortex-like pattern as shown in (\fig{fig:bound_defects}a and \fig{fig:Binding_mech} blue dotted lines).
Using the linearity of Stokes equation, the flow field associated with this configuration is obtained by superposition of the flow fields of each individual $+1/2$ defect that is known analytically~\cite{giomi2014defect}:

\begin{align}
\vec{v}=
\begin{pmatrix}
-\left | \zeta \right |\left( 3 (r_2-r_1)+r_1 \cos (2 \psi_1)- r_2 \cos (2 \psi_2)\right)\\ 
-\left | \zeta \right |\left( r_1 \sin (2 \psi_1)- r_2 \sin (2 \psi_2)\right)
\end{pmatrix},\nonumber
\end{align}
where, $r_{1,2}$ and $\psi_{1,2}$ are the  polar coordinates around defect 1 or 2 at the coordinates $(a,0)$ and $(-a,0)$ with defect orientation $\psi_1=\pi$ and $\psi_2=0$.
In the vicinity of the defect core, the total active flow is non-zero (\fig{fig:bound_defects}a red streamlines), and indeed this tends to destroy the vortex-like configurations as $+1/2$ defect rotate and move away from each other in flow-tumbling regime ($|\nu|<1$).
However, just above the transition to the flow-aligning regime ($|\nu|=1$), the director field tends to align with an angle (Leslie angle) $\theta_L = \frac{1}{2}\sin^{-1}(1/\nu)$ \cite{aigouy2010cell} to the principal axis of flow deformation
\begin{equation}
\theta_P=\frac{1}{2}\tan ^{-1}\left({\frac{2E_{xy}}{E_{xx}-E_{yy}}}\right),\nonumber
\end{equation}
defined based on the rate of strain tensor $\tens{E}$~\cite{voituriez2005spontaneous}.
As such, calculating the reconfiguration of the director field in response to the activity-induced flow of the vortex-like structure (See SM for a detailed calculation) \cite{suppmat} shows that as $\nu\rightarrow 1$, i.e., Leslie angle $\rightarrow \pi/4$, flow gradients result in a preferred reorientation of the director field back into the vortex-like configuration (\fig{fig:Binding_mech}a green dotted lines). Thus $+1/2$ defects move closer together due to an effective attractive force from the activity, which eventually is balanced by the elastic force (See \cite{suppmat}).
This indicates a positive feedback between the active flows and the reorientation response of the director field, which can stabilize the vortex structures and explains the formation of stable bound defect pairs. 

Interestingly, this mechanism of synergy between active flows and flow-aligning behaviour predicts that in an extensile system, where $+1/2$ defects self-propel along their comet-shaped head (\fig{fig:bound_defects}d), stable bound pairs of $+1/2$ defects can form when director tends to align with a Leslie angle to the perpendicular to the principal axis of flow deformations, i.e., for a negative flow-aligning parameter ($\nu < 0$). 
Our calculation of the defect-induced flows and the consecutive reconfiguration of the director field shows that in this case - unlike for contractile case - stable bound defect pairs form when $+1/2$ defects come together head-on, forming aster-like configuration (\fig{fig:Binding_mech}b.)
To test this prediction in our simulations, we consider $\lambda=-0.7$ ($\nu=-1.2$) and increase the extensile activity (\fig{fig:bound_defects}f).
At small activities, bend instabilities are followed by $\pm 1/2$ defects unbinding and active turbulence generation. 
Increasing activities further, we observe the formation of bound $+1/2$ defect pairs in the form of aster-like structures. 
Further increase in activity breaks down these stable bound defects, and active turbulence is recovered. 
As in the contractile case, no aster-like configuration is obtained in a flow-tumbling regime. 
While previous theoretical analyses have predicted the possible existence of such defect configuration due to the active torques acting on $+1/2$ defects~\cite{shankar2018defect}, these results show that the stabilization is a consequence of flow-aligning behaviour of director in response to active flows.
\begin{figure}[b] 
    \centering
    \includegraphics[width=0.45\textwidth]{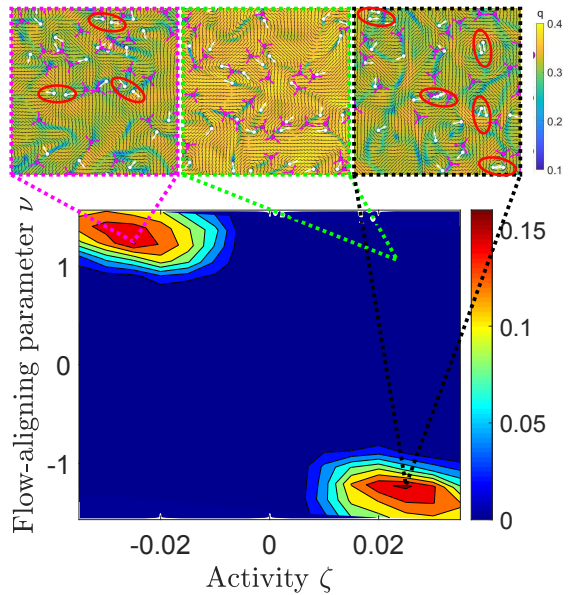}
    \caption{
    Stability diagram of bound defect density (shown in colormap) in the activity flow-aligning parameter ($\zeta-\nu$) phase space. {The colomap in the inset figures indicates the magnitude of nematic order $q$.}
}
\label{fig:stability}
\end{figure}

To more quantitatively characterise the formation of stable bound $+1/2$ defect pairs, we measure the time $+1/2$ defects spend near each other and calculate the density of these bound defect occurrences.
If two self-propelled $+1/2$ defects are for a prolonged time $(t>t^*)$ next to each other $(r<r^*)$ in the configuration as shown in \fig{fig:bound_defects} b,e, they are considered to be bound and stable.
Here $t^*$ is a long time ($t^*=2000$ in simulation units) and $r^*=10$ is a small distance corresponding to the size of vortex- and aster-like configurations.
This is because even in the active turbulence regime, there is always a finite chance of defects being temporarily near each other due to the chaotic nature of $+1/2$ defect motion. A further spatio-temporal characterization of the bound defect life-time and distance between the defects is provided in \cite{suppmat}.

The results of these measurements are shown in \fig{fig:bound_Defects_time} for varying activities and flow-aligning parameters, and clearly indicate that for defects to bound for long times, the system needs to be active, but that the probability of finding a bound defect pair becomes lower eventually with increasing activity (\fig{fig:bound_Defects_time}a). 
This suggests that the $+1/2$ defects need their self-propulsion to form stable vortex- and aster-like structures, but that for higher activities, the more chaotic flows make it easier for the bound defects to break apart. This breaking up is caused by approaching nearby $-1/2$ defects as shown in {\fig{fig:bound_defects}g and}  Movies 1 and 2 in \cite{suppmat}.

\begin{figure}[h]
\centering
\includegraphics[width=0.45\textwidth]{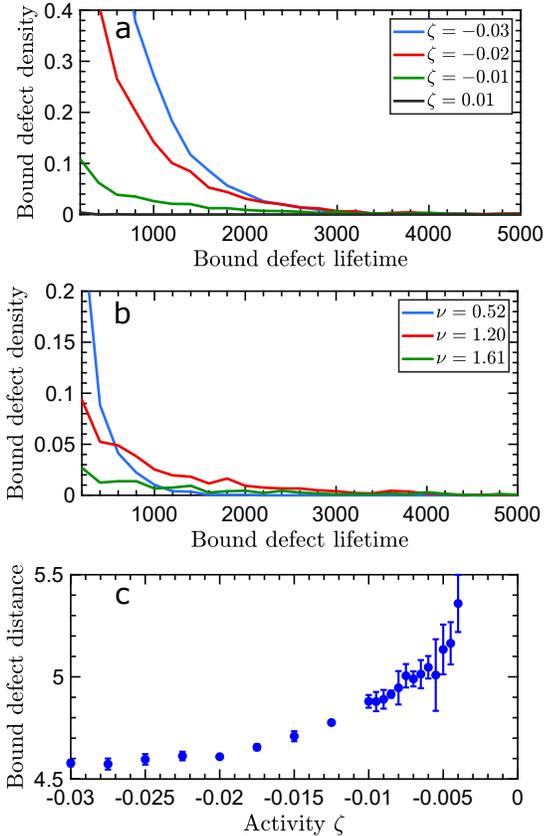}
\caption{{Histogram of the bounded defect lifetime for (a) different activities when $\nu>1$ or (b) for different flow aligning parameters when $\zeta=-0.01$. (c) The average distance between two bound $+1/2$ defects as a function of activity for $\nu>1$.}}
\label{fig:histogram}
\end{figure}

Secondly, we find that the bound defects appear when the flow-aligning parameter approaches $\nu\rightarrow \left | 1 \right |$, above which bound defect density drops again. This is consistent with our analytical predictions that show increasing $\nu$ above 1, makes the director field align stronger with the principal axis of flow deformation, i.e. a smaller Leslie angle, resulting in deviation from reconfiguration shown in \fig{fig:Binding_mech} (See \cite{suppmat} for a detailed calculation) \cite{suppmat}. Eventually for large $\nu$, this results in preferred reconfiguration which renders the bound defect state unstable (Fig.2a,b in the SM) \cite{suppmat}. This explains why eventually the bound defect density drops by increasing flow-aligning parameter.
The results of varying activity and flow-aligning parameters are summarised in the stability diagram (\fig{fig:stability})
showing that similar behaviour of bound defect pair formation is observed for contractile and extensile active nematics in the flow-aligning regime upon changing the sign of the flow-aligning parameter. {A further spatio-temporal characterization of the bound defect life-time and distance between the defects is provided in \fig{fig:histogram}.}
{Measuring the lifetime of two defects near each other shows that this bound defect lifetime has an almost exponential decay (\fig{fig:histogram}a,b). When the activity or the flow aligning parameter is too low, no long-lived bound defects are detected. Interestingly, when the flow aligning parameter becomes too large $\nu$ (similar to the activity) the probability of finding a bound defect goes down for any possible lifetime, and the bound defect density starts to decrease.}

{Moreover, measuring the average distance between bound defects shows that when the activity becomes smaller, the distance between the $+1/2$ defects becomes larger, which suggests that the activity indeed drives the defects close towards each other, which is stabilized with the repulsive elastic force. Increasing the activity results in an eventually plateauing of the distance between $+1/2$ defects due to the repulsive elastic force (\fig{fig:histogram}c).}

{Additionally, characterising the formation of virtual full-integer nematic defects observed here, contribute to the emerging roles of full integer defects in biological systems \textit{e.g.} in formation of protrusions during animal regeneration~\cite{maroudas2020topological} and escape to the third-dimension in fast-moving bacteria~\cite{meacock2020bacteria}. More specifically, Maroudas {\it et al.}~\cite{maroudas2020topological} show that nematic organisation of actin filaments on the surface of the regenerating animal {\it Hydra} are important determinants of the shape of the animal. More importantly, they show that the full-integer defects demarcate sites of head, foot and tentacles in the animal. It was previously shown that in active nematic shells of microtubule-motor protein assemblies pairs of half-integer nematic defects are formed and follow periodic trajectories~\cite{keber2014topology,metselaar2019topology}. Based on the results of our study, we believe that in the regenerating Hydra the full-integer defects are sites where pairs of half-integer defects stabilise into a full-integer structure allowing for protrusions to form. Even closer to the predictions of our work, a recent experiment by Meacock {\it et al.}~\cite{meacock2020bacteria} has revealed that in colonies of fast-moving bacteria pairs of active $+1/2$ defects are able to come close to each other and form virtual full-integer defects. The results are reproduced by a discrete model of self-propelled rods interacting via volume exclusion. The experiments show that these sites of stable bound defects are the points where bacteria escape to the third dimension. Interestingly, and in-line with our predictions here, it is observed that this phenomenon only happens for fast-moving (more active, larger $\zeta$) and longer (larger flow-aligning parameter $\nu$) bacteria compared to the wild-types.
These two recent experiments on different biological systems further indicate the potential importance of full-integer defect formation in active systems.} 

The results presented here, demonstrate a new feature of topological defects in monolayers of active nematics. Moreover, identifying the regions of phase-space where such stable defect binding can be achieved can provide guidelines to design and engineer active nematic systems with more complex topological states

{In addition to these experimental implications, the recent theoretical advances have taken important steps in constructing generic formalisms for multi-defect dynamics in 2D active nematics~\cite{shankar2019hydrodynamics,vafa2020multi} in the limit of compressible and hydrodynamically over-damped systems. Incorporating shear-alignment effects within such a framework and investigating bound defect states would be another interesting step towards constructing generic models of multi-defect dynamics in active systems.}

Finally, very recently, experiments and theories have shown the emergence and complex dynamics of three-dimensional topological defect lines in active nematics~\cite{shendruk2018twist,duclos2020topological,binysh2020three}. Based on the findings presented in this study, we conjecture that similar physics could govern the binding of self-propelled disclination lines in 3D active nematics to form new topologically protected structures.

\section{acknowledgements}

K.T. acknowledges funding from the European Union’s Horizon 2020 research and innovation programme under the Marie Sklodowska-Curie Grant Agreement No. 722497 (LubISS). A. D. acknowledges support from the Novo Nordisk Foundation (grant No. NNF18SA0035142), Villum Fonden (Grant no. 29476), Danish Council for Independent Research, Natural Sciences (DFF-117155-1001), and funding from the European Union’s Horizon 2020 research and innovation program under the Marie Sklodowska-Curie grant agreement No. 847523 (INTERACTIONS).

\bibliographystyle{apsrev4-1}
\bibliography{Bound}

\begin{thebibliography}{72}%
\makeatletter
\providecommand \@ifxundefined [1]{%
 \@ifx{#1\undefined}
}%
\providecommand \@ifnum [1]{%
 \ifnum #1\expandafter \@firstoftwo
 \else \expandafter \@secondoftwo
 \fi
}%
\providecommand \@ifx [1]{%
 \ifx #1\expandafter \@firstoftwo
 \else \expandafter \@secondoftwo
 \fi
}%
\providecommand \natexlab [1]{#1}%
\providecommand \enquote  [1]{``#1''}%
\providecommand \bibnamefont  [1]{#1}%
\providecommand \bibfnamefont [1]{#1}%
\providecommand \citenamefont [1]{#1}%
\providecommand \href@noop [0]{\@secondoftwo}%
\providecommand \href [0]{\begingroup \@sanitize@url \@href}%
\providecommand \@href[1]{\@@startlink{#1}\@@href}%
\providecommand \@@href[1]{\endgroup#1\@@endlink}%
\providecommand \@sanitize@url [0]{\catcode `\\12\catcode `\$12\catcode
  `\&12\catcode `\#12\catcode `\^12\catcode `\_12\catcode `\%12\relax}%
\providecommand \@@startlink[1]{}%
\providecommand \@@endlink[0]{}%
\providecommand \url  [0]{\begingroup\@sanitize@url \@url }%
\providecommand \@url [1]{\endgroup\@href {#1}{\urlprefix }}%
\providecommand \urlprefix  [0]{URL }%
\providecommand \Eprint [0]{\href }%
\providecommand \doibase [0]{http://dx.doi.org/}%
\providecommand \selectlanguage [0]{\@gobble}%
\providecommand \bibinfo  [0]{\@secondoftwo}%
\providecommand \bibfield  [0]{\@secondoftwo}%
\providecommand \translation [1]{[#1]}%
\providecommand \BibitemOpen [0]{}%
\providecommand \bibitemStop [0]{}%
\providecommand \bibitemNoStop [0]{.\EOS\space}%
\providecommand \EOS [0]{\spacefactor3000\relax}%
\providecommand \BibitemShut  [1]{\csname bibitem#1\endcsname}%
\let\auto@bib@innerbib\@empty
\bibitem [{\citenamefont {McLaughlin}\ and\ \citenamefont
  {Scott}(1978)}]{mclaughlin1978perturbation}%
  \BibitemOpen
  \bibfield  {author} {\bibinfo {author} {\bibfnamefont {D.~W.}\ \bibnamefont
  {McLaughlin}}\ and\ \bibinfo {author} {\bibfnamefont {A.~C.}\ \bibnamefont
  {Scott}},\ }\href@noop {} {\bibfield  {journal} {\bibinfo  {journal}
  {Physical Review A}\ }\textbf {\bibinfo {volume} {18}},\ \bibinfo {pages}
  {1652} (\bibinfo {year} {1978})}\BibitemShut {NoStop}%
\bibitem [{\citenamefont {Sakai}\ \emph {et~al.}(1993)\citenamefont {Sakai},
  \citenamefont {Bodin},\ and\ \citenamefont {Pedersen}}]{sakai1993fluxons}%
  \BibitemOpen
  \bibfield  {author} {\bibinfo {author} {\bibfnamefont {S.}~\bibnamefont
  {Sakai}}, \bibinfo {author} {\bibfnamefont {P.}~\bibnamefont {Bodin}}, \ and\
  \bibinfo {author} {\bibfnamefont {N.~F.}\ \bibnamefont {Pedersen}},\
  }\href@noop {} {\bibfield  {journal} {\bibinfo  {journal} {Journal of applied
  physics}\ }\textbf {\bibinfo {volume} {73}},\ \bibinfo {pages} {2411}
  (\bibinfo {year} {1993})}\BibitemShut {NoStop}%
\bibitem [{\citenamefont {Wambaugh}\ \emph {et~al.}(1999)\citenamefont
  {Wambaugh}, \citenamefont {Reichhardt}, \citenamefont {Olson}, \citenamefont
  {Marchesoni},\ and\ \citenamefont {Nori}}]{wambaugh1999superconducting}%
  \BibitemOpen
  \bibfield  {author} {\bibinfo {author} {\bibfnamefont {J.}~\bibnamefont
  {Wambaugh}}, \bibinfo {author} {\bibfnamefont {C.}~\bibnamefont
  {Reichhardt}}, \bibinfo {author} {\bibfnamefont {C.}~\bibnamefont {Olson}},
  \bibinfo {author} {\bibfnamefont {F.}~\bibnamefont {Marchesoni}}, \ and\
  \bibinfo {author} {\bibfnamefont {F.}~\bibnamefont {Nori}},\ }\href@noop {}
  {\bibfield  {journal} {\bibinfo  {journal} {Physical Review Letters}\
  }\textbf {\bibinfo {volume} {83}},\ \bibinfo {pages} {5106} (\bibinfo {year}
  {1999})}\BibitemShut {NoStop}%
\bibitem [{\citenamefont {Salomaa}\ and\ \citenamefont
  {Volovik}(1987)}]{salomaa1987quantized}%
  \BibitemOpen
  \bibfield  {author} {\bibinfo {author} {\bibfnamefont {M.}~\bibnamefont
  {Salomaa}}\ and\ \bibinfo {author} {\bibfnamefont {G.}~\bibnamefont
  {Volovik}},\ }\href@noop {} {\bibfield  {journal} {\bibinfo  {journal}
  {Reviews of modern physics}\ }\textbf {\bibinfo {volume} {59}},\ \bibinfo
  {pages} {533} (\bibinfo {year} {1987})}\BibitemShut {NoStop}%
\bibitem [{\citenamefont {Schwarz}(1988)}]{schwarz1988three}%
  \BibitemOpen
  \bibfield  {author} {\bibinfo {author} {\bibfnamefont {K.}~\bibnamefont
  {Schwarz}},\ }\href@noop {} {\bibfield  {journal} {\bibinfo  {journal}
  {Physical Review B}\ }\textbf {\bibinfo {volume} {38}},\ \bibinfo {pages}
  {2398} (\bibinfo {year} {1988})}\BibitemShut {NoStop}%
\bibitem [{\citenamefont {Zwierlein}\ \emph {et~al.}(2005)\citenamefont
  {Zwierlein}, \citenamefont {Abo-Shaeer}, \citenamefont {Schirotzek},
  \citenamefont {Schunck},\ and\ \citenamefont
  {Ketterle}}]{zwierlein2005vortices}%
  \BibitemOpen
  \bibfield  {author} {\bibinfo {author} {\bibfnamefont {M.~W.}\ \bibnamefont
  {Zwierlein}}, \bibinfo {author} {\bibfnamefont {J.~R.}\ \bibnamefont
  {Abo-Shaeer}}, \bibinfo {author} {\bibfnamefont {A.}~\bibnamefont
  {Schirotzek}}, \bibinfo {author} {\bibfnamefont {C.~H.}\ \bibnamefont
  {Schunck}}, \ and\ \bibinfo {author} {\bibfnamefont {W.}~\bibnamefont
  {Ketterle}},\ }\href@noop {} {\bibfield  {journal} {\bibinfo  {journal}
  {Nature}\ }\textbf {\bibinfo {volume} {435}},\ \bibinfo {pages} {1047}
  (\bibinfo {year} {2005})}\BibitemShut {NoStop}%
\bibitem [{\citenamefont {De~Gennes}\ and\ \citenamefont
  {Prost}(1993)}]{DeGennes}%
  \BibitemOpen
  \bibfield  {author} {\bibinfo {author} {\bibfnamefont {P.-G.}\ \bibnamefont
  {De~Gennes}}\ and\ \bibinfo {author} {\bibfnamefont {J.}~\bibnamefont
  {Prost}},\ }\href@noop {} {\emph {\bibinfo {title} {The physics of liquid
  crystals}}},\ Vol.~\bibinfo {volume} {83}\ (\bibinfo  {publisher} {Oxford
  university press},\ \bibinfo {year} {1993})\BibitemShut {NoStop}%
\bibitem [{\citenamefont {Fert}\ \emph {et~al.}(2013)\citenamefont {Fert},
  \citenamefont {Cros},\ and\ \citenamefont {Sampaio}}]{fert2013skyrmions}%
  \BibitemOpen
  \bibfield  {author} {\bibinfo {author} {\bibfnamefont {A.}~\bibnamefont
  {Fert}}, \bibinfo {author} {\bibfnamefont {V.}~\bibnamefont {Cros}}, \ and\
  \bibinfo {author} {\bibfnamefont {J.}~\bibnamefont {Sampaio}},\ }\href@noop
  {} {\bibfield  {journal} {\bibinfo  {journal} {Nature nanotechnology}\
  }\textbf {\bibinfo {volume} {8}},\ \bibinfo {pages} {152} (\bibinfo {year}
  {2013})}\BibitemShut {NoStop}%
\bibitem [{\citenamefont {Nagaosa}\ and\ \citenamefont
  {Tokura}(2013)}]{nagaosa2013topological}%
  \BibitemOpen
  \bibfield  {author} {\bibinfo {author} {\bibfnamefont {N.}~\bibnamefont
  {Nagaosa}}\ and\ \bibinfo {author} {\bibfnamefont {Y.}~\bibnamefont
  {Tokura}},\ }\href@noop {} {\bibfield  {journal} {\bibinfo  {journal} {Nature
  nanotechnology}\ }\textbf {\bibinfo {volume} {8}},\ \bibinfo {pages} {899}
  (\bibinfo {year} {2013})}\BibitemShut {NoStop}%
\bibitem [{\citenamefont {Wright}\ and\ \citenamefont
  {Mermin}(1989)}]{wright1989crystalline}%
  \BibitemOpen
  \bibfield  {author} {\bibinfo {author} {\bibfnamefont {D.~C.}\ \bibnamefont
  {Wright}}\ and\ \bibinfo {author} {\bibfnamefont {N.~D.}\ \bibnamefont
  {Mermin}},\ }\href@noop {} {\bibfield  {journal} {\bibinfo  {journal}
  {Reviews of Modern physics}\ }\textbf {\bibinfo {volume} {61}},\ \bibinfo
  {pages} {385} (\bibinfo {year} {1989})}\BibitemShut {NoStop}%
\bibitem [{\citenamefont {Livshits}\ \emph {et~al.}(2017)\citenamefont
  {Livshits}, \citenamefont {Shani-Zerbib}, \citenamefont {Maroudas-Sacks},
  \citenamefont {Braun},\ and\ \citenamefont {Keren}}]{livshits2017structural}%
  \BibitemOpen
  \bibfield  {author} {\bibinfo {author} {\bibfnamefont {A.}~\bibnamefont
  {Livshits}}, \bibinfo {author} {\bibfnamefont {L.}~\bibnamefont
  {Shani-Zerbib}}, \bibinfo {author} {\bibfnamefont {Y.}~\bibnamefont
  {Maroudas-Sacks}}, \bibinfo {author} {\bibfnamefont {E.}~\bibnamefont
  {Braun}}, \ and\ \bibinfo {author} {\bibfnamefont {K.}~\bibnamefont
  {Keren}},\ }\href@noop {} {\bibfield  {journal} {\bibinfo  {journal} {Cell
  reports}\ }\textbf {\bibinfo {volume} {18}},\ \bibinfo {pages} {1410}
  (\bibinfo {year} {2017})}\BibitemShut {NoStop}%
\bibitem [{\citenamefont {Maroudas-Sacks}\ \emph {et~al.}(2020)\citenamefont
  {Maroudas-Sacks}, \citenamefont {Garion}, \citenamefont {Shani-Zerbib},
  \citenamefont {Livshits}, \citenamefont {Braun},\ and\ \citenamefont
  {Keren}}]{maroudas2020topological}%
  \BibitemOpen
  \bibfield  {author} {\bibinfo {author} {\bibfnamefont {Y.}~\bibnamefont
  {Maroudas-Sacks}}, \bibinfo {author} {\bibfnamefont {L.}~\bibnamefont
  {Garion}}, \bibinfo {author} {\bibfnamefont {L.}~\bibnamefont
  {Shani-Zerbib}}, \bibinfo {author} {\bibfnamefont {A.}~\bibnamefont
  {Livshits}}, \bibinfo {author} {\bibfnamefont {E.}~\bibnamefont {Braun}}, \
  and\ \bibinfo {author} {\bibfnamefont {K.}~\bibnamefont {Keren}},\
  }\href@noop {} {\bibfield  {journal} {\bibinfo  {journal} {bioRxiv}\ }
  (\bibinfo {year} {2020})}\BibitemShut {NoStop}%
\bibitem [{\citenamefont {Saw}\ \emph {et~al.}(2017)\citenamefont {Saw},
  \citenamefont {Doostmohammadi}, \citenamefont {Nier}, \citenamefont
  {Kocgozlu}, \citenamefont {Thampi}, \citenamefont {Toyama}, \citenamefont
  {Marcq}, \citenamefont {Lim}, \citenamefont {Yeomans},\ and\ \citenamefont
  {Ladoux}}]{Saw2017}%
  \BibitemOpen
  \bibfield  {author} {\bibinfo {author} {\bibfnamefont {T.~B.}\ \bibnamefont
  {Saw}}, \bibinfo {author} {\bibfnamefont {A.}~\bibnamefont {Doostmohammadi}},
  \bibinfo {author} {\bibfnamefont {V.}~\bibnamefont {Nier}}, \bibinfo {author}
  {\bibfnamefont {L.}~\bibnamefont {Kocgozlu}}, \bibinfo {author}
  {\bibfnamefont {S.}~\bibnamefont {Thampi}}, \bibinfo {author} {\bibfnamefont
  {Y.}~\bibnamefont {Toyama}}, \bibinfo {author} {\bibfnamefont
  {P.}~\bibnamefont {Marcq}}, \bibinfo {author} {\bibfnamefont {C.~T.}\
  \bibnamefont {Lim}}, \bibinfo {author} {\bibfnamefont {J.~M.}\ \bibnamefont
  {Yeomans}}, \ and\ \bibinfo {author} {\bibfnamefont {B.}~\bibnamefont
  {Ladoux}},\ }\href@noop {} {\bibfield  {journal} {\bibinfo  {journal}
  {Nature}\ }\textbf {\bibinfo {volume} {544}},\ \bibinfo {pages} {212}
  (\bibinfo {year} {2017})}\BibitemShut {NoStop}%
\bibitem [{\citenamefont {Yaman}\ \emph {et~al.}(2019)\citenamefont {Yaman},
  \citenamefont {Demir}, \citenamefont {Vetter},\ and\ \citenamefont
  {Kocabas}}]{yaman2019}%
  \BibitemOpen
  \bibfield  {author} {\bibinfo {author} {\bibfnamefont {Y.~I.}\ \bibnamefont
  {Yaman}}, \bibinfo {author} {\bibfnamefont {E.}~\bibnamefont {Demir}},
  \bibinfo {author} {\bibfnamefont {R.}~\bibnamefont {Vetter}}, \ and\ \bibinfo
  {author} {\bibfnamefont {A.}~\bibnamefont {Kocabas}},\ }\href@noop {}
  {\bibfield  {journal} {\bibinfo  {journal} {Nature communications}\ }\textbf
  {\bibinfo {volume} {10}},\ \bibinfo {pages} {1} (\bibinfo {year}
  {2019})}\BibitemShut {NoStop}%
\bibitem [{\citenamefont {Kawaguchi}\ \emph {et~al.}(2017)\citenamefont
  {Kawaguchi}, \citenamefont {Kageyama},\ and\ \citenamefont
  {Sano}}]{kawaguchi2017topological}%
  \BibitemOpen
  \bibfield  {author} {\bibinfo {author} {\bibfnamefont {K.}~\bibnamefont
  {Kawaguchi}}, \bibinfo {author} {\bibfnamefont {R.}~\bibnamefont {Kageyama}},
  \ and\ \bibinfo {author} {\bibfnamefont {M.}~\bibnamefont {Sano}},\
  }\href@noop {} {\bibfield  {journal} {\bibinfo  {journal} {Nature}\ }\textbf
  {\bibinfo {volume} {545}},\ \bibinfo {pages} {327} (\bibinfo {year}
  {2017})}\BibitemShut {NoStop}%
\bibitem [{\citenamefont {Doostmohammadi}\ \emph {et~al.}(2018)\citenamefont
  {Doostmohammadi}, \citenamefont {Ign{\'e}s-Mullol}, \citenamefont {Yeomans},\
  and\ \citenamefont {Sagu{\'e}s}}]{doostmohammadi2018active}%
  \BibitemOpen
  \bibfield  {author} {\bibinfo {author} {\bibfnamefont {A.}~\bibnamefont
  {Doostmohammadi}}, \bibinfo {author} {\bibfnamefont {J.}~\bibnamefont
  {Ign{\'e}s-Mullol}}, \bibinfo {author} {\bibfnamefont {J.~M.}\ \bibnamefont
  {Yeomans}}, \ and\ \bibinfo {author} {\bibfnamefont {F.}~\bibnamefont
  {Sagu{\'e}s}},\ }\href@noop {} {\bibfield  {journal} {\bibinfo  {journal}
  {Nature communications}\ }\textbf {\bibinfo {volume} {9}},\ \bibinfo {pages}
  {1} (\bibinfo {year} {2018})}\BibitemShut {NoStop}%
\bibitem [{\citenamefont {Ramaswamy}(2010)}]{ramaswamy2010mechanics}%
  \BibitemOpen
  \bibfield  {author} {\bibinfo {author} {\bibfnamefont {S.}~\bibnamefont
  {Ramaswamy}},\ }\href@noop {} {\bibfield  {journal} {\bibinfo  {journal}
  {Annu. Rev. Condens. Matter Phys.}\ }\textbf {\bibinfo {volume} {1}},\
  \bibinfo {pages} {323} (\bibinfo {year} {2010})}\BibitemShut {NoStop}%
\bibitem [{\citenamefont {Marchetti}\ \emph {et~al.}(2013)\citenamefont
  {Marchetti}, \citenamefont {Joanny}, \citenamefont {Ramaswamy}, \citenamefont
  {Liverpool}, \citenamefont {Prost}, \citenamefont {Rao},\ and\ \citenamefont
  {Simha}}]{Marchetti13}%
  \BibitemOpen
  \bibfield  {author} {\bibinfo {author} {\bibfnamefont {M.~C.}\ \bibnamefont
  {Marchetti}}, \bibinfo {author} {\bibfnamefont {J.~F.}\ \bibnamefont
  {Joanny}}, \bibinfo {author} {\bibfnamefont {S.}~\bibnamefont {Ramaswamy}},
  \bibinfo {author} {\bibfnamefont {T.~B.}\ \bibnamefont {Liverpool}}, \bibinfo
  {author} {\bibfnamefont {J.}~\bibnamefont {Prost}}, \bibinfo {author}
  {\bibfnamefont {M.}~\bibnamefont {Rao}}, \ and\ \bibinfo {author}
  {\bibfnamefont {R.~A.}\ \bibnamefont {Simha}},\ }\href@noop {} {\bibfield
  {journal} {\bibinfo  {journal} {Rev. Mod. Phys.}\ }\textbf {\bibinfo {volume}
  {85}},\ \bibinfo {pages} {1143} (\bibinfo {year} {2013})}\BibitemShut
  {NoStop}%
\bibitem [{\citenamefont {Bechinger}\ \emph {et~al.}(2016)\citenamefont
  {Bechinger}, \citenamefont {Di~Leonardo}, \citenamefont {L\"owen},
  \citenamefont {Reichhardt}, \citenamefont {Volpe},\ and\ \citenamefont
  {Volpe}}]{Bechinger2016}%
  \BibitemOpen
  \bibfield  {author} {\bibinfo {author} {\bibfnamefont {C.}~\bibnamefont
  {Bechinger}}, \bibinfo {author} {\bibfnamefont {R.}~\bibnamefont
  {Di~Leonardo}}, \bibinfo {author} {\bibfnamefont {H.}~\bibnamefont
  {L\"owen}}, \bibinfo {author} {\bibfnamefont {C.}~\bibnamefont {Reichhardt}},
  \bibinfo {author} {\bibfnamefont {G.}~\bibnamefont {Volpe}}, \ and\ \bibinfo
  {author} {\bibfnamefont {G.}~\bibnamefont {Volpe}},\ }\href@noop {}
  {\bibfield  {journal} {\bibinfo  {journal} {Rev. Mod. Phys.}\ }\textbf
  {\bibinfo {volume} {88}},\ \bibinfo {pages} {045006} (\bibinfo {year}
  {2016})}\BibitemShut {NoStop}%
\bibitem [{\citenamefont {J{\"u}licher}\ \emph {et~al.}(2018)\citenamefont
  {J{\"u}licher}, \citenamefont {Grill},\ and\ \citenamefont
  {Salbreux}}]{julicher2018hydrodynamic}%
  \BibitemOpen
  \bibfield  {author} {\bibinfo {author} {\bibfnamefont {F.}~\bibnamefont
  {J{\"u}licher}}, \bibinfo {author} {\bibfnamefont {S.~W.}\ \bibnamefont
  {Grill}}, \ and\ \bibinfo {author} {\bibfnamefont {G.}~\bibnamefont
  {Salbreux}},\ }\href@noop {} {\bibfield  {journal} {\bibinfo  {journal}
  {Reports on Progress in Physics}\ }\textbf {\bibinfo {volume} {81}},\
  \bibinfo {pages} {076601} (\bibinfo {year} {2018})}\BibitemShut {NoStop}%
\bibitem [{\citenamefont {Sanchez}\ \emph {et~al.}(2012)\citenamefont
  {Sanchez}, \citenamefont {Chen}, \citenamefont {DeCamp}, \citenamefont
  {Heymann},\ and\ \citenamefont {Dogic}}]{Sanchez2012}%
  \BibitemOpen
  \bibfield  {author} {\bibinfo {author} {\bibfnamefont {T.}~\bibnamefont
  {Sanchez}}, \bibinfo {author} {\bibfnamefont {D.~T.~N.}\ \bibnamefont
  {Chen}}, \bibinfo {author} {\bibfnamefont {S.~J.}\ \bibnamefont {DeCamp}},
  \bibinfo {author} {\bibfnamefont {M.}~\bibnamefont {Heymann}}, \ and\
  \bibinfo {author} {\bibfnamefont {Z.}~\bibnamefont {Dogic}},\ }\href@noop {}
  {\bibfield  {journal} {\bibinfo  {journal} {Nature}\ }\textbf {\bibinfo
  {volume} {491}},\ \bibinfo {pages} {431} (\bibinfo {year}
  {2012})}\BibitemShut {NoStop}%
\bibitem [{\citenamefont {Giomi}\ \emph {et~al.}(2014)\citenamefont {Giomi},
  \citenamefont {Bowick}, \citenamefont {Mishra}, \citenamefont {Sknepnek},\
  and\ \citenamefont {Cristina~Marchetti}}]{giomi2014defect}%
  \BibitemOpen
  \bibfield  {author} {\bibinfo {author} {\bibfnamefont {L.}~\bibnamefont
  {Giomi}}, \bibinfo {author} {\bibfnamefont {M.~J.}\ \bibnamefont {Bowick}},
  \bibinfo {author} {\bibfnamefont {P.}~\bibnamefont {Mishra}}, \bibinfo
  {author} {\bibfnamefont {R.}~\bibnamefont {Sknepnek}}, \ and\ \bibinfo
  {author} {\bibfnamefont {M.}~\bibnamefont {Cristina~Marchetti}},\ }\href@noop
  {} {\bibfield  {journal} {\bibinfo  {journal} {Philosophical Transactions of
  the Royal Society A: Mathematical, Physical and Engineering Sciences}\
  }\textbf {\bibinfo {volume} {372}},\ \bibinfo {pages} {20130365} (\bibinfo
  {year} {2014})}\BibitemShut {NoStop}%
\bibitem [{\citenamefont {Kumar}\ \emph {et~al.}(2018)\citenamefont {Kumar},
  \citenamefont {Zhang}, \citenamefont {de~Pablo},\ and\ \citenamefont
  {Gardel}}]{kumar2018tunable}%
  \BibitemOpen
  \bibfield  {author} {\bibinfo {author} {\bibfnamefont {N.}~\bibnamefont
  {Kumar}}, \bibinfo {author} {\bibfnamefont {R.}~\bibnamefont {Zhang}},
  \bibinfo {author} {\bibfnamefont {J.~J.}\ \bibnamefont {de~Pablo}}, \ and\
  \bibinfo {author} {\bibfnamefont {M.~L.}\ \bibnamefont {Gardel}},\
  }\href@noop {} {\bibfield  {journal} {\bibinfo  {journal} {Science Advances}\
  }\textbf {\bibinfo {volume} {4}},\ \bibinfo {pages} {eaat7779} (\bibinfo
  {year} {2018})}\BibitemShut {NoStop}%
\bibitem [{\citenamefont {Li}\ \emph {et~al.}(2019)\citenamefont {Li},
  \citenamefont {Shi}, \citenamefont {Huang}, \citenamefont {Chen},
  \citenamefont {Xiao}, \citenamefont {Liu}, \citenamefont {Chat{\'e}},\ and\
  \citenamefont {Zhang}}]{Li19}%
  \BibitemOpen
  \bibfield  {author} {\bibinfo {author} {\bibfnamefont {H.}~\bibnamefont
  {Li}}, \bibinfo {author} {\bibfnamefont {X.-q.}\ \bibnamefont {Shi}},
  \bibinfo {author} {\bibfnamefont {M.}~\bibnamefont {Huang}}, \bibinfo
  {author} {\bibfnamefont {X.}~\bibnamefont {Chen}}, \bibinfo {author}
  {\bibfnamefont {M.}~\bibnamefont {Xiao}}, \bibinfo {author} {\bibfnamefont
  {C.}~\bibnamefont {Liu}}, \bibinfo {author} {\bibfnamefont {H.}~\bibnamefont
  {Chat{\'e}}}, \ and\ \bibinfo {author} {\bibfnamefont {H.}~\bibnamefont
  {Zhang}},\ }\href@noop {} {\bibfield  {journal} {\bibinfo  {journal} {PNAS}\
  }\textbf {\bibinfo {volume} {116}},\ \bibinfo {pages} {777} (\bibinfo {year}
  {2019})}\BibitemShut {NoStop}%
\bibitem [{\citenamefont {Shankar}\ \emph {et~al.}(2018)\citenamefont
  {Shankar}, \citenamefont {Ramaswamy}, \citenamefont {Marchetti},\ and\
  \citenamefont {Bowick}}]{shankar2018defect}%
  \BibitemOpen
  \bibfield  {author} {\bibinfo {author} {\bibfnamefont {S.}~\bibnamefont
  {Shankar}}, \bibinfo {author} {\bibfnamefont {S.}~\bibnamefont {Ramaswamy}},
  \bibinfo {author} {\bibfnamefont {M.~C.}\ \bibnamefont {Marchetti}}, \ and\
  \bibinfo {author} {\bibfnamefont {M.~J.}\ \bibnamefont {Bowick}},\
  }\href@noop {} {\bibfield  {journal} {\bibinfo  {journal} {Physical review
  letters}\ }\textbf {\bibinfo {volume} {121}},\ \bibinfo {pages} {108002}
  (\bibinfo {year} {2018})}\BibitemShut {NoStop}%
\bibitem [{\citenamefont {DeCamp}\ \emph {et~al.}(2015)\citenamefont {DeCamp},
  \citenamefont {Redner}, \citenamefont {Baskaran}, \citenamefont {Hagan},\
  and\ \citenamefont {Dogic}}]{decamp2015orientational}%
  \BibitemOpen
  \bibfield  {author} {\bibinfo {author} {\bibfnamefont {S.~J.}\ \bibnamefont
  {DeCamp}}, \bibinfo {author} {\bibfnamefont {G.~S.}\ \bibnamefont {Redner}},
  \bibinfo {author} {\bibfnamefont {A.}~\bibnamefont {Baskaran}}, \bibinfo
  {author} {\bibfnamefont {M.~F.}\ \bibnamefont {Hagan}}, \ and\ \bibinfo
  {author} {\bibfnamefont {Z.}~\bibnamefont {Dogic}},\ }\href@noop {}
  {\bibfield  {journal} {\bibinfo  {journal} {Nat. Mat.}\ }\textbf {\bibinfo
  {volume} {14}},\ \bibinfo {pages} {1110} (\bibinfo {year}
  {2015})}\BibitemShut {NoStop}%
\bibitem [{\citenamefont {Thijssen}\ \emph {et~al.}(2020)\citenamefont
  {Thijssen}, \citenamefont {Nejad},\ and\ \citenamefont
  {Yeomans}}]{thijssen2020large}%
  \BibitemOpen
  \bibfield  {author} {\bibinfo {author} {\bibfnamefont {K.}~\bibnamefont
  {Thijssen}}, \bibinfo {author} {\bibfnamefont {M.~R.}\ \bibnamefont {Nejad}},
  \ and\ \bibinfo {author} {\bibfnamefont {J.~M.}\ \bibnamefont {Yeomans}},\
  }\href@noop {} {\bibfield  {journal} {\bibinfo  {journal} {arXiv:2005.01164}\
  } (\bibinfo {year} {2020})}\BibitemShut {NoStop}%
\bibitem [{\citenamefont {Pearce}\ \emph {et~al.}(2020)\citenamefont {Pearce},
  \citenamefont {Nambisan}, \citenamefont {Ellis}, \citenamefont {Dogic},
  \citenamefont {Fernandez-Nieves},\ and\ \citenamefont
  {Giomi}}]{pearce2020scale}%
  \BibitemOpen
  \bibfield  {author} {\bibinfo {author} {\bibfnamefont {D.}~\bibnamefont
  {Pearce}}, \bibinfo {author} {\bibfnamefont {J.}~\bibnamefont {Nambisan}},
  \bibinfo {author} {\bibfnamefont {P.}~\bibnamefont {Ellis}}, \bibinfo
  {author} {\bibfnamefont {Z.}~\bibnamefont {Dogic}}, \bibinfo {author}
  {\bibfnamefont {A.}~\bibnamefont {Fernandez-Nieves}}, \ and\ \bibinfo
  {author} {\bibfnamefont {L.}~\bibnamefont {Giomi}},\ }\href@noop {}
  {\bibfield  {journal} {\bibinfo  {journal} {arXiv preprint arXiv:2004.13704}\
  } (\bibinfo {year} {2020})}\BibitemShut {NoStop}%
\bibitem [{\citenamefont {Putzig}\ \emph {et~al.}(2016)\citenamefont {Putzig},
  \citenamefont {Redner}, \citenamefont {Baskaran},\ and\ \citenamefont
  {Baskaran}}]{putzig2016instabilities}%
  \BibitemOpen
  \bibfield  {author} {\bibinfo {author} {\bibfnamefont {E.}~\bibnamefont
  {Putzig}}, \bibinfo {author} {\bibfnamefont {G.~S.}\ \bibnamefont {Redner}},
  \bibinfo {author} {\bibfnamefont {A.}~\bibnamefont {Baskaran}}, \ and\
  \bibinfo {author} {\bibfnamefont {A.}~\bibnamefont {Baskaran}},\ }\href@noop
  {} {\bibfield  {journal} {\bibinfo  {journal} {Soft matter}\ }\textbf
  {\bibinfo {volume} {12}},\ \bibinfo {pages} {3854} (\bibinfo {year}
  {2016})}\BibitemShut {NoStop}%
\bibitem [{\citenamefont {Shankar}\ and\ \citenamefont
  {Marchetti}(2019)}]{shankar2019hydrodynamics}%
  \BibitemOpen
  \bibfield  {author} {\bibinfo {author} {\bibfnamefont {S.}~\bibnamefont
  {Shankar}}\ and\ \bibinfo {author} {\bibfnamefont {M.~C.}\ \bibnamefont
  {Marchetti}},\ }\href@noop {} {\bibfield  {journal} {\bibinfo  {journal}
  {Physical Review X}\ }\textbf {\bibinfo {volume} {9}},\ \bibinfo {pages}
  {041047} (\bibinfo {year} {2019})}\BibitemShut {NoStop}%
\bibitem [{\citenamefont {Doostmohammadi}\ \emph
  {et~al.}(2016{\natexlab{a}})\citenamefont {Doostmohammadi}, \citenamefont
  {Adamer}, \citenamefont {Thampi},\ and\ \citenamefont
  {Yeomans}}]{doostmohammadi2016stabilization}%
  \BibitemOpen
  \bibfield  {author} {\bibinfo {author} {\bibfnamefont {A.}~\bibnamefont
  {Doostmohammadi}}, \bibinfo {author} {\bibfnamefont {M.~F.}\ \bibnamefont
  {Adamer}}, \bibinfo {author} {\bibfnamefont {S.~P.}\ \bibnamefont {Thampi}},
  \ and\ \bibinfo {author} {\bibfnamefont {J.~M.}\ \bibnamefont {Yeomans}},\
  }\href@noop {} {\bibfield  {journal} {\bibinfo  {journal} {Nat. Comm.}\
  }\textbf {\bibinfo {volume} {7}},\ \bibinfo {pages} {10557} (\bibinfo {year}
  {2016}{\natexlab{a}})}\BibitemShut {NoStop}%
\bibitem [{\citenamefont {Oza}\ and\ \citenamefont
  {Dunkel}(2016)}]{oza2016antipolar}%
  \BibitemOpen
  \bibfield  {author} {\bibinfo {author} {\bibfnamefont {A.~U.}\ \bibnamefont
  {Oza}}\ and\ \bibinfo {author} {\bibfnamefont {J.}~\bibnamefont {Dunkel}},\
  }\href@noop {} {\bibfield  {journal} {\bibinfo  {journal} {New Journal of
  Physics}\ }\textbf {\bibinfo {volume} {18}},\ \bibinfo {pages} {093006}
  (\bibinfo {year} {2016})}\BibitemShut {NoStop}%
\bibitem [{\citenamefont {Doostmohammadi}\ \emph
  {et~al.}(2016{\natexlab{b}})\citenamefont {Doostmohammadi}, \citenamefont
  {Thampi},\ and\ \citenamefont {Yeomans}}]{doostmohammadi2016defect}%
  \BibitemOpen
  \bibfield  {author} {\bibinfo {author} {\bibfnamefont {A.}~\bibnamefont
  {Doostmohammadi}}, \bibinfo {author} {\bibfnamefont {S.~P.}\ \bibnamefont
  {Thampi}}, \ and\ \bibinfo {author} {\bibfnamefont {J.~M.}\ \bibnamefont
  {Yeomans}},\ }\href@noop {} {\bibfield  {journal} {\bibinfo  {journal} {Phys.
  Rev. Lett.}\ }\textbf {\bibinfo {volume} {117}},\ \bibinfo {pages} {048102}
  (\bibinfo {year} {2016}{\natexlab{b}})}\BibitemShut {NoStop}%
\bibitem [{\citenamefont {Metselaar}\ \emph {et~al.}(2019)\citenamefont
  {Metselaar}, \citenamefont {Yeomans},\ and\ \citenamefont
  {Doostmohammadi}}]{metselaar2019topology}%
  \BibitemOpen
  \bibfield  {author} {\bibinfo {author} {\bibfnamefont {L.}~\bibnamefont
  {Metselaar}}, \bibinfo {author} {\bibfnamefont {J.~M.}\ \bibnamefont
  {Yeomans}}, \ and\ \bibinfo {author} {\bibfnamefont {A.}~\bibnamefont
  {Doostmohammadi}},\ }\href@noop {} {\bibfield  {journal} {\bibinfo  {journal}
  {Physical Review Letters}\ }\textbf {\bibinfo {volume} {123}},\ \bibinfo
  {pages} {208001} (\bibinfo {year} {2019})}\BibitemShut {NoStop}%
\bibitem [{\citenamefont {Dell’Arciprete}\ \emph {et~al.}(2018)\citenamefont
  {Dell’Arciprete}, \citenamefont {Blow}, \citenamefont {Brown},
  \citenamefont {Farrell}, \citenamefont {Lintuvuori}, \citenamefont {McVey},
  \citenamefont {Marenduzzo},\ and\ \citenamefont {Poon}}]{dell2018growing}%
  \BibitemOpen
  \bibfield  {author} {\bibinfo {author} {\bibfnamefont {D.}~\bibnamefont
  {Dell’Arciprete}}, \bibinfo {author} {\bibfnamefont {M.}~\bibnamefont
  {Blow}}, \bibinfo {author} {\bibfnamefont {A.}~\bibnamefont {Brown}},
  \bibinfo {author} {\bibfnamefont {F.}~\bibnamefont {Farrell}}, \bibinfo
  {author} {\bibfnamefont {J.~S.}\ \bibnamefont {Lintuvuori}}, \bibinfo
  {author} {\bibfnamefont {A.}~\bibnamefont {McVey}}, \bibinfo {author}
  {\bibfnamefont {D.}~\bibnamefont {Marenduzzo}}, \ and\ \bibinfo {author}
  {\bibfnamefont {W.~C.}\ \bibnamefont {Poon}},\ }\href@noop {} {\bibfield
  {journal} {\bibinfo  {journal} {Nature communications}\ }\textbf {\bibinfo
  {volume} {9}},\ \bibinfo {pages} {1} (\bibinfo {year} {2018})}\BibitemShut
  {NoStop}%
\bibitem [{\citenamefont {Morris}\ and\ \citenamefont
  {Rao}(2019)}]{morris2019active}%
  \BibitemOpen
  \bibfield  {author} {\bibinfo {author} {\bibfnamefont {R.~G.}\ \bibnamefont
  {Morris}}\ and\ \bibinfo {author} {\bibfnamefont {M.}~\bibnamefont {Rao}},\
  }\href@noop {} {\bibfield  {journal} {\bibinfo  {journal} {Physical Review
  E}\ }\textbf {\bibinfo {volume} {100}},\ \bibinfo {pages} {022413} (\bibinfo
  {year} {2019})}\BibitemShut {NoStop}%
\bibitem [{\citenamefont {Meyer}(1973)}]{meyer1973existence}%
  \BibitemOpen
  \bibfield  {author} {\bibinfo {author} {\bibfnamefont {R.~B.}\ \bibnamefont
  {Meyer}},\ }\href@noop {} {\bibfield  {journal} {\bibinfo  {journal} {The
  Philosophical Magazine: A Journal of Theoretical Experimental and Applied
  Physics}\ }\textbf {\bibinfo {volume} {27}},\ \bibinfo {pages} {405}
  (\bibinfo {year} {1973})}\BibitemShut {NoStop}%
\bibitem [{\citenamefont {Wensink}\ \emph {et~al.}(2012)\citenamefont
  {Wensink}, \citenamefont {Dunkel}, \citenamefont {Heidenreich}, \citenamefont
  {Drescher}, \citenamefont {Goldstein}, \citenamefont {L{\"o}wen},\ and\
  \citenamefont {Yeomans}}]{Wensink2012}%
  \BibitemOpen
  \bibfield  {author} {\bibinfo {author} {\bibfnamefont {H.}~\bibnamefont
  {Wensink}}, \bibinfo {author} {\bibfnamefont {J.}~\bibnamefont {Dunkel}},
  \bibinfo {author} {\bibfnamefont {S.}~\bibnamefont {Heidenreich}}, \bibinfo
  {author} {\bibfnamefont {K.}~\bibnamefont {Drescher}}, \bibinfo {author}
  {\bibfnamefont {R.~E.}\ \bibnamefont {Goldstein}}, \bibinfo {author}
  {\bibfnamefont {H.}~\bibnamefont {L{\"o}wen}}, \ and\ \bibinfo {author}
  {\bibfnamefont {J.~M.}\ \bibnamefont {Yeomans}},\ }\href@noop {} {\bibfield
  {journal} {\bibinfo  {journal} {PNAS}\ }\textbf {\bibinfo {volume} {109}},\
  \bibinfo {pages} {14308} (\bibinfo {year} {2012})}\BibitemShut {NoStop}%
\bibitem [{\citenamefont {Lee}\ and\ \citenamefont
  {Kardar}(2001)}]{lee2001macroscopic}%
  \BibitemOpen
  \bibfield  {author} {\bibinfo {author} {\bibfnamefont {H.~Y.}\ \bibnamefont
  {Lee}}\ and\ \bibinfo {author} {\bibfnamefont {M.}~\bibnamefont {Kardar}},\
  }\href@noop {} {\bibfield  {journal} {\bibinfo  {journal} {Physical Review
  E}\ }\textbf {\bibinfo {volume} {64}},\ \bibinfo {pages} {056113} (\bibinfo
  {year} {2001})}\BibitemShut {NoStop}%
\bibitem [{\citenamefont {Kruse}\ \emph {et~al.}(2004)\citenamefont {Kruse},
  \citenamefont {Joanny}, \citenamefont {J{\"u}licher}, \citenamefont {Prost},\
  and\ \citenamefont {Sekimoto}}]{kruse2004asters}%
  \BibitemOpen
  \bibfield  {author} {\bibinfo {author} {\bibfnamefont {K.}~\bibnamefont
  {Kruse}}, \bibinfo {author} {\bibfnamefont {J.-F.}\ \bibnamefont {Joanny}},
  \bibinfo {author} {\bibfnamefont {F.}~\bibnamefont {J{\"u}licher}}, \bibinfo
  {author} {\bibfnamefont {J.}~\bibnamefont {Prost}}, \ and\ \bibinfo {author}
  {\bibfnamefont {K.}~\bibnamefont {Sekimoto}},\ }\href@noop {} {\bibfield
  {journal} {\bibinfo  {journal} {Physical review letters}\ }\textbf {\bibinfo
  {volume} {92}},\ \bibinfo {pages} {078101} (\bibinfo {year}
  {2004})}\BibitemShut {NoStop}%
\bibitem [{\citenamefont {Husain}\ and\ \citenamefont
  {Rao}(2017)}]{husain2017emergent}%
  \BibitemOpen
  \bibfield  {author} {\bibinfo {author} {\bibfnamefont {K.}~\bibnamefont
  {Husain}}\ and\ \bibinfo {author} {\bibfnamefont {M.}~\bibnamefont {Rao}},\
  }\href@noop {} {\bibfield  {journal} {\bibinfo  {journal} {Physical review
  letters}\ }\textbf {\bibinfo {volume} {118}},\ \bibinfo {pages} {078104}
  (\bibinfo {year} {2017})}\BibitemShut {NoStop}%
\bibitem [{\citenamefont {Endresen}\ \emph {et~al.}(2019)\citenamefont
  {Endresen}, \citenamefont {Kim},\ and\ \citenamefont
  {Serra}}]{endresen2019topological}%
  \BibitemOpen
  \bibfield  {author} {\bibinfo {author} {\bibfnamefont {K.~D.}\ \bibnamefont
  {Endresen}}, \bibinfo {author} {\bibfnamefont {M.}~\bibnamefont {Kim}}, \
  and\ \bibinfo {author} {\bibfnamefont {F.}~\bibnamefont {Serra}},\
  }\href@noop {} {\bibfield  {journal} {\bibinfo  {journal} {arXiv preprint
  arXiv:1912.03271}\ } (\bibinfo {year} {2019})}\BibitemShut {NoStop}%
\bibitem [{\citenamefont {Duclos}\ \emph {et~al.}(2017)\citenamefont {Duclos},
  \citenamefont {Erlenk{\"a}mper}, \citenamefont {Joanny},\ and\ \citenamefont
  {Silberzan}}]{duclos2017topological}%
  \BibitemOpen
  \bibfield  {author} {\bibinfo {author} {\bibfnamefont {G.}~\bibnamefont
  {Duclos}}, \bibinfo {author} {\bibfnamefont {C.}~\bibnamefont
  {Erlenk{\"a}mper}}, \bibinfo {author} {\bibfnamefont {J.-F.}\ \bibnamefont
  {Joanny}}, \ and\ \bibinfo {author} {\bibfnamefont {P.}~\bibnamefont
  {Silberzan}},\ }\href@noop {} {\bibfield  {journal} {\bibinfo  {journal}
  {Nature Physics}\ }\textbf {\bibinfo {volume} {13}},\ \bibinfo {pages} {58}
  (\bibinfo {year} {2017})}\BibitemShut {NoStop}%
\bibitem [{\citenamefont {Opathalage}\ \emph {et~al.}(2019)\citenamefont
  {Opathalage}, \citenamefont {Norton}, \citenamefont {Juniper}, \citenamefont
  {Langeslay}, \citenamefont {Aghvami}, \citenamefont {Fraden},\ and\
  \citenamefont {Dogic}}]{opathalage2019self}%
  \BibitemOpen
  \bibfield  {author} {\bibinfo {author} {\bibfnamefont {A.}~\bibnamefont
  {Opathalage}}, \bibinfo {author} {\bibfnamefont {M.~M.}\ \bibnamefont
  {Norton}}, \bibinfo {author} {\bibfnamefont {M.~P.}\ \bibnamefont {Juniper}},
  \bibinfo {author} {\bibfnamefont {B.}~\bibnamefont {Langeslay}}, \bibinfo
  {author} {\bibfnamefont {S.~A.}\ \bibnamefont {Aghvami}}, \bibinfo {author}
  {\bibfnamefont {S.}~\bibnamefont {Fraden}}, \ and\ \bibinfo {author}
  {\bibfnamefont {Z.}~\bibnamefont {Dogic}},\ }\href@noop {} {\bibfield
  {journal} {\bibinfo  {journal} {Proceedings of the National Academy of
  Sciences}\ }\textbf {\bibinfo {volume} {116}},\ \bibinfo {pages} {4788}
  (\bibinfo {year} {2019})}\BibitemShut {NoStop}%
\bibitem [{\citenamefont {Guillamat}\ \emph {et~al.}(2020)\citenamefont
  {Guillamat}, \citenamefont {Blanch-Mercader}, \citenamefont {Kruse},\ and\
  \citenamefont {Roux}}]{guillamat2020integer}%
  \BibitemOpen
  \bibfield  {author} {\bibinfo {author} {\bibfnamefont {P.}~\bibnamefont
  {Guillamat}}, \bibinfo {author} {\bibfnamefont {C.}~\bibnamefont
  {Blanch-Mercader}}, \bibinfo {author} {\bibfnamefont {K.}~\bibnamefont
  {Kruse}}, \ and\ \bibinfo {author} {\bibfnamefont {A.}~\bibnamefont {Roux}},\
  }\href@noop {} {\bibfield  {journal} {\bibinfo  {journal} {bioRxiv}\ }
  (\bibinfo {year} {2020})}\BibitemShut {NoStop}%
\bibitem [{\citenamefont {Turiv}\ \emph {et~al.}(2020)\citenamefont {Turiv},
  \citenamefont {Krieger}, \citenamefont {Babakhanova}, \citenamefont {Yu},
  \citenamefont {Shiyanovskii}, \citenamefont {Wei}, \citenamefont {Kim},\ and\
  \citenamefont {Lavrentovich}}]{turiv2020topology}%
  \BibitemOpen
  \bibfield  {author} {\bibinfo {author} {\bibfnamefont {T.}~\bibnamefont
  {Turiv}}, \bibinfo {author} {\bibfnamefont {J.}~\bibnamefont {Krieger}},
  \bibinfo {author} {\bibfnamefont {G.}~\bibnamefont {Babakhanova}}, \bibinfo
  {author} {\bibfnamefont {H.}~\bibnamefont {Yu}}, \bibinfo {author}
  {\bibfnamefont {S.~V.}\ \bibnamefont {Shiyanovskii}}, \bibinfo {author}
  {\bibfnamefont {Q.-H.}\ \bibnamefont {Wei}}, \bibinfo {author} {\bibfnamefont
  {M.-H.}\ \bibnamefont {Kim}}, \ and\ \bibinfo {author} {\bibfnamefont
  {O.~D.}\ \bibnamefont {Lavrentovich}},\ }\href@noop {} {\bibfield  {journal}
  {\bibinfo  {journal} {Science Advances}\ }\textbf {\bibinfo {volume} {6}},\
  \bibinfo {pages} {eaaz6485} (\bibinfo {year} {2020})}\BibitemShut {NoStop}%
\bibitem [{\citenamefont {Rivas}\ \emph {et~al.}(2019)\citenamefont {Rivas},
  \citenamefont {Shendruk}, \citenamefont {Henry}, \citenamefont {Reich},\ and\
  \citenamefont {Leheny}}]{rivas2019driven}%
  \BibitemOpen
  \bibfield  {author} {\bibinfo {author} {\bibfnamefont {D.~P.}\ \bibnamefont
  {Rivas}}, \bibinfo {author} {\bibfnamefont {T.~N.}\ \bibnamefont {Shendruk}},
  \bibinfo {author} {\bibfnamefont {R.~R.}\ \bibnamefont {Henry}}, \bibinfo
  {author} {\bibfnamefont {D.~H.}\ \bibnamefont {Reich}}, \ and\ \bibinfo
  {author} {\bibfnamefont {R.~L.}\ \bibnamefont {Leheny}},\ }\href@noop {}
  {\bibfield  {journal} {\bibinfo  {journal} {arXiv preprint arXiv:1910.14456}\
  } (\bibinfo {year} {2019})}\BibitemShut {NoStop}%
\bibitem [{\citenamefont {Hardo{\"u}in}\ \emph {et~al.}(2019)\citenamefont
  {Hardo{\"u}in}, \citenamefont {Hughes}, \citenamefont {Doostmohammadi},
  \citenamefont {Laurent}, \citenamefont {Lopez-Leon}, \citenamefont {Yeomans},
  \citenamefont {Ign{\'e}s-Mullol},\ and\ \citenamefont
  {Sagu{\'e}s}}]{hardouin2019reconfigurable}%
  \BibitemOpen
  \bibfield  {author} {\bibinfo {author} {\bibfnamefont {J.}~\bibnamefont
  {Hardo{\"u}in}}, \bibinfo {author} {\bibfnamefont {R.}~\bibnamefont
  {Hughes}}, \bibinfo {author} {\bibfnamefont {A.}~\bibnamefont
  {Doostmohammadi}}, \bibinfo {author} {\bibfnamefont {J.}~\bibnamefont
  {Laurent}}, \bibinfo {author} {\bibfnamefont {T.}~\bibnamefont {Lopez-Leon}},
  \bibinfo {author} {\bibfnamefont {J.~M.}\ \bibnamefont {Yeomans}}, \bibinfo
  {author} {\bibfnamefont {J.}~\bibnamefont {Ign{\'e}s-Mullol}}, \ and\
  \bibinfo {author} {\bibfnamefont {F.}~\bibnamefont {Sagu{\'e}s}},\
  }\href@noop {} {\bibfield  {journal} {\bibinfo  {journal} {Communications
  Physics}\ }\textbf {\bibinfo {volume} {2}},\ \bibinfo {pages} {1} (\bibinfo
  {year} {2019})}\BibitemShut {NoStop}%
\bibitem [{\citenamefont {Zhang}\ \emph {et~al.}(2018)\citenamefont {Zhang},
  \citenamefont {Kumar}, \citenamefont {Ross}, \citenamefont {Gardel},\ and\
  \citenamefont {De~Pablo}}]{zhang2018interplay}%
  \BibitemOpen
  \bibfield  {author} {\bibinfo {author} {\bibfnamefont {R.}~\bibnamefont
  {Zhang}}, \bibinfo {author} {\bibfnamefont {N.}~\bibnamefont {Kumar}},
  \bibinfo {author} {\bibfnamefont {J.~L.}\ \bibnamefont {Ross}}, \bibinfo
  {author} {\bibfnamefont {M.~L.}\ \bibnamefont {Gardel}}, \ and\ \bibinfo
  {author} {\bibfnamefont {J.~J.}\ \bibnamefont {De~Pablo}},\ }\href@noop {}
  {\bibfield  {journal} {\bibinfo  {journal} {Proceedings of the National
  Academy of Sciences}\ }\textbf {\bibinfo {volume} {115}},\ \bibinfo {pages}
  {E124} (\bibinfo {year} {2018})}\BibitemShut {NoStop}%
\bibitem [{\citenamefont {Volfson}\ \emph {et~al.}(2008)\citenamefont
  {Volfson}, \citenamefont {Cookson}, \citenamefont {Hasty},\ and\
  \citenamefont {Tsimring}}]{volfson2008biomechanical}%
  \BibitemOpen
  \bibfield  {author} {\bibinfo {author} {\bibfnamefont {D.}~\bibnamefont
  {Volfson}}, \bibinfo {author} {\bibfnamefont {S.}~\bibnamefont {Cookson}},
  \bibinfo {author} {\bibfnamefont {J.}~\bibnamefont {Hasty}}, \ and\ \bibinfo
  {author} {\bibfnamefont {L.~S.}\ \bibnamefont {Tsimring}},\ }\href@noop {}
  {\bibfield  {journal} {\bibinfo  {journal} {Proceedings of the National
  Academy of Sciences}\ }\textbf {\bibinfo {volume} {105}},\ \bibinfo {pages}
  {15346} (\bibinfo {year} {2008})}\BibitemShut {NoStop}%
\bibitem [{\citenamefont {Zhang}\ \emph {et~al.}(2010)\citenamefont {Zhang},
  \citenamefont {Be’er}, \citenamefont {Florin},\ and\ \citenamefont
  {Swinney}}]{zhang2010collective}%
  \BibitemOpen
  \bibfield  {author} {\bibinfo {author} {\bibfnamefont {H.-P.}\ \bibnamefont
  {Zhang}}, \bibinfo {author} {\bibfnamefont {A.}~\bibnamefont {Be’er}},
  \bibinfo {author} {\bibfnamefont {E.-L.}\ \bibnamefont {Florin}}, \ and\
  \bibinfo {author} {\bibfnamefont {H.~L.}\ \bibnamefont {Swinney}},\
  }\href@noop {} {\bibfield  {journal} {\bibinfo  {journal} {Proceedings of the
  National Academy of Sciences}\ }\textbf {\bibinfo {volume} {107}},\ \bibinfo
  {pages} {13626} (\bibinfo {year} {2010})}\BibitemShut {NoStop}%
\bibitem [{\citenamefont {Beroz}\ \emph {et~al.}(2018)\citenamefont {Beroz},
  \citenamefont {Yan}, \citenamefont {Meir}, \citenamefont {Sabass},
  \citenamefont {Stone}, \citenamefont {Bassler},\ and\ \citenamefont
  {Wingreen}}]{beroz2018verticalization}%
  \BibitemOpen
  \bibfield  {author} {\bibinfo {author} {\bibfnamefont {F.}~\bibnamefont
  {Beroz}}, \bibinfo {author} {\bibfnamefont {J.}~\bibnamefont {Yan}}, \bibinfo
  {author} {\bibfnamefont {Y.}~\bibnamefont {Meir}}, \bibinfo {author}
  {\bibfnamefont {B.}~\bibnamefont {Sabass}}, \bibinfo {author} {\bibfnamefont
  {H.~A.}\ \bibnamefont {Stone}}, \bibinfo {author} {\bibfnamefont {B.~L.}\
  \bibnamefont {Bassler}}, \ and\ \bibinfo {author} {\bibfnamefont {N.~S.}\
  \bibnamefont {Wingreen}},\ }\href@noop {} {\bibfield  {journal} {\bibinfo
  {journal} {Nature physics}\ }\textbf {\bibinfo {volume} {14}},\ \bibinfo
  {pages} {954} (\bibinfo {year} {2018})}\BibitemShut {NoStop}%
\bibitem [{\citenamefont {Duclos}\ \emph {et~al.}(2014)\citenamefont {Duclos},
  \citenamefont {Garcia}, \citenamefont {Yevick},\ and\ \citenamefont
  {Silberzan}}]{duclos2014perfect}%
  \BibitemOpen
  \bibfield  {author} {\bibinfo {author} {\bibfnamefont {G.}~\bibnamefont
  {Duclos}}, \bibinfo {author} {\bibfnamefont {S.}~\bibnamefont {Garcia}},
  \bibinfo {author} {\bibfnamefont {H.}~\bibnamefont {Yevick}}, \ and\ \bibinfo
  {author} {\bibfnamefont {P.}~\bibnamefont {Silberzan}},\ }\href@noop {}
  {\bibfield  {journal} {\bibinfo  {journal} {Soft matter}\ }\textbf {\bibinfo
  {volume} {10}},\ \bibinfo {pages} {2346} (\bibinfo {year}
  {2014})}\BibitemShut {NoStop}%
\bibitem [{sup()}]{suppmat}%
  \BibitemOpen
  \href@noop {} {}\bibinfo {note} {See Supplemental Material at [URL will be
  inserted by publisher] for details.}\BibitemShut {Stop}%
\bibitem [{\citenamefont {Beris}\ and\ \citenamefont
  {Edwards}(1994)}]{BerisBook}%
  \BibitemOpen
  \bibfield  {author} {\bibinfo {author} {\bibfnamefont {A.~N.}\ \bibnamefont
  {Beris}}\ and\ \bibinfo {author} {\bibfnamefont {B.~J.}\ \bibnamefont
  {Edwards}},\ }\href@noop {} {\emph {\bibinfo {title} {{T}hermodynamics of
  {F}lowing {S}ystems}}}\ (\bibinfo  {publisher} {Oxford University Press},\
  \bibinfo {year} {1994})\BibitemShut {NoStop}%
\bibitem [{\citenamefont {Aigouy}\ \emph {et~al.}(2010)\citenamefont {Aigouy},
  \citenamefont {Farhadifar}, \citenamefont {Staple}, \citenamefont {Sagner},
  \citenamefont {R{\"o}per}, \citenamefont {J{\"u}licher},\ and\ \citenamefont
  {Eaton}}]{aigouy2010cell}%
  \BibitemOpen
  \bibfield  {author} {\bibinfo {author} {\bibfnamefont {B.}~\bibnamefont
  {Aigouy}}, \bibinfo {author} {\bibfnamefont {R.}~\bibnamefont {Farhadifar}},
  \bibinfo {author} {\bibfnamefont {D.~B.}\ \bibnamefont {Staple}}, \bibinfo
  {author} {\bibfnamefont {A.}~\bibnamefont {Sagner}}, \bibinfo {author}
  {\bibfnamefont {J.-C.}\ \bibnamefont {R{\"o}per}}, \bibinfo {author}
  {\bibfnamefont {F.}~\bibnamefont {J{\"u}licher}}, \ and\ \bibinfo {author}
  {\bibfnamefont {S.}~\bibnamefont {Eaton}},\ }\href@noop {} {\bibfield
  {journal} {\bibinfo  {journal} {Cell}\ }\textbf {\bibinfo {volume} {142}},\
  \bibinfo {pages} {773} (\bibinfo {year} {2010})}\BibitemShut {NoStop}%
\bibitem [{\citenamefont {Duclos}\ \emph {et~al.}(2018)\citenamefont {Duclos},
  \citenamefont {Blanch-Mercader}, \citenamefont {Yashunsky}, \citenamefont
  {Salbreux}, \citenamefont {Joanny}, \citenamefont {Prost},\ and\
  \citenamefont {Silberzan}}]{duclos2018spontaneous}%
  \BibitemOpen
  \bibfield  {author} {\bibinfo {author} {\bibfnamefont {G.}~\bibnamefont
  {Duclos}}, \bibinfo {author} {\bibfnamefont {C.}~\bibnamefont
  {Blanch-Mercader}}, \bibinfo {author} {\bibfnamefont {V.}~\bibnamefont
  {Yashunsky}}, \bibinfo {author} {\bibfnamefont {G.}~\bibnamefont {Salbreux}},
  \bibinfo {author} {\bibfnamefont {J.-F.}\ \bibnamefont {Joanny}}, \bibinfo
  {author} {\bibfnamefont {J.}~\bibnamefont {Prost}}, \ and\ \bibinfo {author}
  {\bibfnamefont {P.}~\bibnamefont {Silberzan}},\ }\href@noop {} {\bibfield
  {journal} {\bibinfo  {journal} {Nature physics}\ }\textbf {\bibinfo {volume}
  {14}},\ \bibinfo {pages} {728} (\bibinfo {year} {2018})}\BibitemShut
  {NoStop}%
\bibitem [{\citenamefont {Guillamat}\ \emph {et~al.}(2017)\citenamefont
  {Guillamat}, \citenamefont {Ign{\'e}s-Mullol},\ and\ \citenamefont
  {Sagu{\'e}s}}]{guillamat2017taming}%
  \BibitemOpen
  \bibfield  {author} {\bibinfo {author} {\bibfnamefont {P.}~\bibnamefont
  {Guillamat}}, \bibinfo {author} {\bibfnamefont {J.}~\bibnamefont
  {Ign{\'e}s-Mullol}}, \ and\ \bibinfo {author} {\bibfnamefont
  {F.}~\bibnamefont {Sagu{\'e}s}},\ }\href@noop {} {\bibfield  {journal}
  {\bibinfo  {journal} {Nature communications}\ }\textbf {\bibinfo {volume}
  {8}},\ \bibinfo {pages} {1} (\bibinfo {year} {2017})}\BibitemShut {NoStop}%
\bibitem [{\citenamefont {Srivastava}\ \emph {et~al.}(2016)\citenamefont
  {Srivastava}, \citenamefont {Mishra},\ and\ \citenamefont
  {Marchetti}}]{srivastava2016negative}%
  \BibitemOpen
  \bibfield  {author} {\bibinfo {author} {\bibfnamefont {P.}~\bibnamefont
  {Srivastava}}, \bibinfo {author} {\bibfnamefont {P.}~\bibnamefont {Mishra}},
  \ and\ \bibinfo {author} {\bibfnamefont {M.~C.}\ \bibnamefont {Marchetti}},\
  }\href@noop {} {\bibfield  {journal} {\bibinfo  {journal} {Soft Matter}\
  }\textbf {\bibinfo {volume} {12}},\ \bibinfo {pages} {8214} (\bibinfo {year}
  {2016})}\BibitemShut {NoStop}%
\bibitem [{\citenamefont {Doostmohammadi}\ \emph {et~al.}(2017)\citenamefont
  {Doostmohammadi}, \citenamefont {Shendruk}, \citenamefont {Thijssen},\ and\
  \citenamefont {Yeomans}}]{doostmohammadi2017onset}%
  \BibitemOpen
  \bibfield  {author} {\bibinfo {author} {\bibfnamefont {A.}~\bibnamefont
  {Doostmohammadi}}, \bibinfo {author} {\bibfnamefont {T.~N.}\ \bibnamefont
  {Shendruk}}, \bibinfo {author} {\bibfnamefont {K.}~\bibnamefont {Thijssen}},
  \ and\ \bibinfo {author} {\bibfnamefont {J.~M.}\ \bibnamefont {Yeomans}},\
  }\href@noop {} {\bibfield  {journal} {\bibinfo  {journal} {Nature
  communications}\ }\textbf {\bibinfo {volume} {8}},\ \bibinfo {pages} {1}
  (\bibinfo {year} {2017})}\BibitemShut {NoStop}%
\bibitem [{\citenamefont {Shendruk}\ \emph {et~al.}(2017)\citenamefont
  {Shendruk}, \citenamefont {Doostmohammadi}, \citenamefont {Thijssen},\ and\
  \citenamefont {Yeomans}}]{shendruk2017dancing}%
  \BibitemOpen
  \bibfield  {author} {\bibinfo {author} {\bibfnamefont {T.~N.}\ \bibnamefont
  {Shendruk}}, \bibinfo {author} {\bibfnamefont {A.}~\bibnamefont
  {Doostmohammadi}}, \bibinfo {author} {\bibfnamefont {K.}~\bibnamefont
  {Thijssen}}, \ and\ \bibinfo {author} {\bibfnamefont {J.~M.}\ \bibnamefont
  {Yeomans}},\ }\href@noop {} {\bibfield  {journal} {\bibinfo  {journal} {Soft
  Matter}\ }\textbf {\bibinfo {volume} {13}},\ \bibinfo {pages} {3853}
  (\bibinfo {year} {2017})}\BibitemShut {NoStop}%
\bibitem [{\citenamefont {Santhosh}\ \emph {et~al.}(2020)\citenamefont
  {Santhosh}, \citenamefont {Nejad}, \citenamefont {Doostmohammadi},
  \citenamefont {Yeomans},\ and\ \citenamefont
  {Thampi}}]{santhosh2020activity}%
  \BibitemOpen
  \bibfield  {author} {\bibinfo {author} {\bibfnamefont {S.}~\bibnamefont
  {Santhosh}}, \bibinfo {author} {\bibfnamefont {M.~R.}\ \bibnamefont {Nejad}},
  \bibinfo {author} {\bibfnamefont {A.}~\bibnamefont {Doostmohammadi}},
  \bibinfo {author} {\bibfnamefont {J.~M.}\ \bibnamefont {Yeomans}}, \ and\
  \bibinfo {author} {\bibfnamefont {S.~P.}\ \bibnamefont {Thampi}},\
  }\href@noop {} {\bibfield  {journal} {\bibinfo  {journal} {Journal of
  Statistical Physics}\ ,\ \bibinfo {pages} {1}} (\bibinfo {year}
  {2020})}\BibitemShut {NoStop}%
\bibitem [{\citenamefont {Chandragiri}\ \emph {et~al.}(2019)\citenamefont
  {Chandragiri}, \citenamefont {Doostmohammadi}, \citenamefont {Yeomans},\ and\
  \citenamefont {Thampi}}]{chandragiri2019active}%
  \BibitemOpen
  \bibfield  {author} {\bibinfo {author} {\bibfnamefont {S.}~\bibnamefont
  {Chandragiri}}, \bibinfo {author} {\bibfnamefont {A.}~\bibnamefont
  {Doostmohammadi}}, \bibinfo {author} {\bibfnamefont {J.~M.}\ \bibnamefont
  {Yeomans}}, \ and\ \bibinfo {author} {\bibfnamefont {S.~P.}\ \bibnamefont
  {Thampi}},\ }\href@noop {} {\bibfield  {journal} {\bibinfo  {journal} {Soft
  matter}\ }\textbf {\bibinfo {volume} {15}},\ \bibinfo {pages} {1597}
  (\bibinfo {year} {2019})}\BibitemShut {NoStop}%
\bibitem [{\citenamefont {Giomi}\ \emph {et~al.}(2013)\citenamefont {Giomi},
  \citenamefont {Bowick}, \citenamefont {Ma},\ and\ \citenamefont
  {Marchetti}}]{giomi2013defect}%
  \BibitemOpen
  \bibfield  {author} {\bibinfo {author} {\bibfnamefont {L.}~\bibnamefont
  {Giomi}}, \bibinfo {author} {\bibfnamefont {M.~J.}\ \bibnamefont {Bowick}},
  \bibinfo {author} {\bibfnamefont {X.}~\bibnamefont {Ma}}, \ and\ \bibinfo
  {author} {\bibfnamefont {M.~C.}\ \bibnamefont {Marchetti}},\ }\href@noop {}
  {\bibfield  {journal} {\bibinfo  {journal} {Physical review letters}\
  }\textbf {\bibinfo {volume} {110}},\ \bibinfo {pages} {228101} (\bibinfo
  {year} {2013})}\BibitemShut {NoStop}%
\bibitem [{\citenamefont {Thampi}\ \emph {et~al.}(2014)\citenamefont {Thampi},
  \citenamefont {Golestanian},\ and\ \citenamefont
  {Yeomans}}]{thampi2014instabilities}%
  \BibitemOpen
  \bibfield  {author} {\bibinfo {author} {\bibfnamefont {S.~P.}\ \bibnamefont
  {Thampi}}, \bibinfo {author} {\bibfnamefont {R.}~\bibnamefont {Golestanian}},
  \ and\ \bibinfo {author} {\bibfnamefont {J.~M.}\ \bibnamefont {Yeomans}},\
  }\href@noop {} {\bibfield  {journal} {\bibinfo  {journal} {EPL (Europhysics
  Letters)}\ }\textbf {\bibinfo {volume} {105}},\ \bibinfo {pages} {18001}
  (\bibinfo {year} {2014})}\BibitemShut {NoStop}%
\bibitem [{\citenamefont {Voituriez}\ \emph {et~al.}(2005)\citenamefont
  {Voituriez}, \citenamefont {Joanny},\ and\ \citenamefont
  {Prost}}]{voituriez2005spontaneous}%
  \BibitemOpen
  \bibfield  {author} {\bibinfo {author} {\bibfnamefont {R.}~\bibnamefont
  {Voituriez}}, \bibinfo {author} {\bibfnamefont {J.-F.}\ \bibnamefont
  {Joanny}}, \ and\ \bibinfo {author} {\bibfnamefont {J.}~\bibnamefont
  {Prost}},\ }\href@noop {} {\bibfield  {journal} {\bibinfo  {journal} {EPL
  (Europhysics Letters)}\ }\textbf {\bibinfo {volume} {70}},\ \bibinfo {pages}
  {404} (\bibinfo {year} {2005})}\BibitemShut {NoStop}%
\bibitem [{\citenamefont {Meacock}\ \emph {et~al.}(2020)\citenamefont
  {Meacock}, \citenamefont {Doostmohammadi}, \citenamefont {Foster},
  \citenamefont {Yeomans},\ and\ \citenamefont {Durham}}]{meacock2020bacteria}%
  \BibitemOpen
  \bibfield  {author} {\bibinfo {author} {\bibfnamefont {O.~J.}\ \bibnamefont
  {Meacock}}, \bibinfo {author} {\bibfnamefont {A.}~\bibnamefont
  {Doostmohammadi}}, \bibinfo {author} {\bibfnamefont {K.~R.}\ \bibnamefont
  {Foster}}, \bibinfo {author} {\bibfnamefont {J.~M.}\ \bibnamefont {Yeomans}},
  \ and\ \bibinfo {author} {\bibfnamefont {W.~M.}\ \bibnamefont {Durham}},\
  }\href@noop {} {\bibfield  {journal} {\bibinfo  {journal} {arXiv preprint
  arXiv:2008.07915}\ } (\bibinfo {year} {2020})}\BibitemShut {NoStop}%
\bibitem [{\citenamefont {Keber}\ \emph {et~al.}(2014)\citenamefont {Keber},
  \citenamefont {Loiseau}, \citenamefont {Sanchez}, \citenamefont {DeCamp},
  \citenamefont {Giomi}, \citenamefont {Bowick}, \citenamefont {Marchetti},
  \citenamefont {Dogic},\ and\ \citenamefont {Bausch}}]{keber2014topology}%
  \BibitemOpen
  \bibfield  {author} {\bibinfo {author} {\bibfnamefont {F.~C.}\ \bibnamefont
  {Keber}}, \bibinfo {author} {\bibfnamefont {E.}~\bibnamefont {Loiseau}},
  \bibinfo {author} {\bibfnamefont {T.}~\bibnamefont {Sanchez}}, \bibinfo
  {author} {\bibfnamefont {S.~J.}\ \bibnamefont {DeCamp}}, \bibinfo {author}
  {\bibfnamefont {L.}~\bibnamefont {Giomi}}, \bibinfo {author} {\bibfnamefont
  {M.~J.}\ \bibnamefont {Bowick}}, \bibinfo {author} {\bibfnamefont {M.~C.}\
  \bibnamefont {Marchetti}}, \bibinfo {author} {\bibfnamefont {Z.}~\bibnamefont
  {Dogic}}, \ and\ \bibinfo {author} {\bibfnamefont {A.~R.}\ \bibnamefont
  {Bausch}},\ }\href@noop {} {\bibfield  {journal} {\bibinfo  {journal}
  {Science}\ }\textbf {\bibinfo {volume} {345}},\ \bibinfo {pages} {1135}
  (\bibinfo {year} {2014})}\BibitemShut {NoStop}%
\bibitem [{\citenamefont {Vafa}\ \emph {et~al.}(2020)\citenamefont {Vafa},
  \citenamefont {Bowick}, \citenamefont {Marchetti},\ and\ \citenamefont
  {Shraiman}}]{vafa2020multi}%
  \BibitemOpen
  \bibfield  {author} {\bibinfo {author} {\bibfnamefont {F.}~\bibnamefont
  {Vafa}}, \bibinfo {author} {\bibfnamefont {M.~J.}\ \bibnamefont {Bowick}},
  \bibinfo {author} {\bibfnamefont {M.~C.}\ \bibnamefont {Marchetti}}, \ and\
  \bibinfo {author} {\bibfnamefont {B.~I.}\ \bibnamefont {Shraiman}},\
  }\href@noop {} {\bibfield  {journal} {\bibinfo  {journal} {arXiv:2007.02947}\
  } (\bibinfo {year} {2020})}\BibitemShut {NoStop}%
\bibitem [{\citenamefont {Shendruk}\ \emph {et~al.}(2018)\citenamefont
  {Shendruk}, \citenamefont {Thijssen}, \citenamefont {Yeomans},\ and\
  \citenamefont {Doostmohammadi}}]{shendruk2018twist}%
  \BibitemOpen
  \bibfield  {author} {\bibinfo {author} {\bibfnamefont {T.~N.}\ \bibnamefont
  {Shendruk}}, \bibinfo {author} {\bibfnamefont {K.}~\bibnamefont {Thijssen}},
  \bibinfo {author} {\bibfnamefont {J.~M.}\ \bibnamefont {Yeomans}}, \ and\
  \bibinfo {author} {\bibfnamefont {A.}~\bibnamefont {Doostmohammadi}},\
  }\href@noop {} {\bibfield  {journal} {\bibinfo  {journal} {Physical Review
  E}\ }\textbf {\bibinfo {volume} {98}},\ \bibinfo {pages} {010601} (\bibinfo
  {year} {2018})}\BibitemShut {NoStop}%
\bibitem [{\citenamefont {Duclos}\ \emph {et~al.}(2020)\citenamefont {Duclos},
  \citenamefont {Adkins}, \citenamefont {Banerjee}, \citenamefont {Peterson},
  \citenamefont {Varghese}, \citenamefont {Kolvin}, \citenamefont {Baskaran},
  \citenamefont {Pelcovits}, \citenamefont {Powers}, \citenamefont {Baskaran}
  \emph {et~al.}}]{duclos2020topological}%
  \BibitemOpen
  \bibfield  {author} {\bibinfo {author} {\bibfnamefont {G.}~\bibnamefont
  {Duclos}}, \bibinfo {author} {\bibfnamefont {R.}~\bibnamefont {Adkins}},
  \bibinfo {author} {\bibfnamefont {D.}~\bibnamefont {Banerjee}}, \bibinfo
  {author} {\bibfnamefont {M.~S.}\ \bibnamefont {Peterson}}, \bibinfo {author}
  {\bibfnamefont {M.}~\bibnamefont {Varghese}}, \bibinfo {author}
  {\bibfnamefont {I.}~\bibnamefont {Kolvin}}, \bibinfo {author} {\bibfnamefont
  {A.}~\bibnamefont {Baskaran}}, \bibinfo {author} {\bibfnamefont {R.~A.}\
  \bibnamefont {Pelcovits}}, \bibinfo {author} {\bibfnamefont {T.~R.}\
  \bibnamefont {Powers}}, \bibinfo {author} {\bibfnamefont {A.}~\bibnamefont
  {Baskaran}},  \emph {et~al.},\ }\href@noop {} {\bibfield  {journal} {\bibinfo
   {journal} {Science}\ }\textbf {\bibinfo {volume} {367}},\ \bibinfo {pages}
  {1120} (\bibinfo {year} {2020})}\BibitemShut {NoStop}%
\bibitem [{\citenamefont {Binysh}\ \emph {et~al.}(2020)\citenamefont {Binysh},
  \citenamefont {Kos}, \citenamefont {{\v{C}}opar}, \citenamefont {Ravnik},\
  and\ \citenamefont {Alexander}}]{binysh2020three}%
  \BibitemOpen
  \bibfield  {author} {\bibinfo {author} {\bibfnamefont {J.}~\bibnamefont
  {Binysh}}, \bibinfo {author} {\bibfnamefont {{\v{Z}}.}~\bibnamefont {Kos}},
  \bibinfo {author} {\bibfnamefont {S.}~\bibnamefont {{\v{C}}opar}}, \bibinfo
  {author} {\bibfnamefont {M.}~\bibnamefont {Ravnik}}, \ and\ \bibinfo {author}
  {\bibfnamefont {G.~P.}\ \bibnamefont {Alexander}},\ }\href@noop {} {\bibfield
   {journal} {\bibinfo  {journal} {Physical Review Letters}\ }\textbf {\bibinfo
  {volume} {124}},\ \bibinfo {pages} {088001} (\bibinfo {year}
  {2020})}\BibitemShut {NoStop}%
\end{thebibliography}%

\end{document}